\documentclass[12pt]{article}
\pdfoutput=1
\usepackage{amsmath, amsthm, amssymb, amsfonts, mathtools, adjustbox}
\usepackage[margin=1in]{geometry}
\usepackage{graphicx}
\usepackage[numbers]{natbib}
\usepackage{hyperref}
\usepackage{bm}
\usepackage[dvipsnames]{xcolor}

\newcommand{\norm}[1]{\left\lVert#1\right\rVert}
\newcommand{\abs}[1]{\left|#1\right|}
\newcommand{\mat}[1]{\begin{bmatrix}#1\end{bmatrix}}
\newcommand{\R}{\mathbb{R}}

\renewcommand{\O}{\mathcal{O}}
\newcommand{\I}{\mathcal{I}}
\newcommand{\id}{\bm{I}}
\DeclareMathOperator{\Cov}{Cov}

\DeclareMathOperator{\blkdiag}{blkdiag}
\DeclareMathOperator{\tr}{tr}
\DeclareMathOperator{\logdet}{logdet}

\newcommand{\pj}[1]{\frac{\partial#1}{\partial\theta_j}}
\newcommand{\prpj}[1]{\bigg(\frac{\partial#1}{\partial\theta_j}\bigg)}
\newcommand{\pk}[1]{\frac{\partial#1}{\partial\theta_k}}
\newcommand{\prpk}[1]{\bigg(\frac{\partial#1}{\partial\theta_k}\bigg)}
\newcommand{\pjk}[1]{\frac{\partial^2#1}{\partial\theta_j\partial\theta_k}}
\newcommand{\prpjk}[1]{\bigg(\frac{\partial^2#1}{\partial\theta_j\partial\theta_k}\bigg)}

\newcommand{\cm}{\bm{\Sigma}} 
\newcommand{\acm}{\widetilde{\bm{\Sigma}}} 
\newcommand{\prms}{\bm{\theta}} 
\newcommand{\mle}{\widehat{\prms}} 
\newcommand{\perm}{\bm{\Pi}} 
\newcommand{\Lam}{\bm{\Lambda}} 

\renewcommand{\a}{\bm{a}}

\newcommand{\p}{\bm{p}}
\newcommand{\w}{\bm{w}}
\newcommand{\x}{\bm{x}}
\newcommand{\y}{\bm{y}}
\newcommand{\z}{\bm{z}}
\renewcommand{\u}{\bm{u}}
\newcommand{\A}{\bm{A}}
\newcommand{\B}{\bm{B}}

\newcommand{\D}{\bm{D}}
\newcommand{\G}{\bm{G}}

\renewcommand{\L}{\bm{L}}
\newcommand{\M}{\bm{M}}
\newcommand{\U}{\bm{U}}
\newcommand{\V}{\bm{V}}
\newcommand{\W}{\bm{W}}
\newcommand{\X}{\bm{X}}
\newcommand{\Y}{\bm{Y}}
\newcommand{\Z}{\bm{Z}}

\title{\bf Scalable Computations for Nonstationary Gaussian Processes}
\author{
  \bf{Paul G. Beckman}\thanks{
    corresponding author \texttt{paul.beckman@cims.nyu.edu}
  } \thanks{
    This material was based upon work supported by the US Department of Energy,
    Office of Science, Office of Advanced Scientific Computing Research (ASCR)
    under Contract DE-AC02-06CH11347
  } \\
  Courant Institute of Mathematical Sciences, New York University \\
  Mathematics and Computer Science Division, Argonne National Laboratory \\[1.5ex]
  \bf{Christopher J. Geoga} \\
  Mathematics and Computer Science Division, Argonne National Laboratory \\
  Department of Statistics, Rutgers University \\[1.5ex]
  \bf{Michael L. Stein} \\
  Department of Statistics, Rutgers University \\
  Department of Statistics, University of Chicago \\[1.5ex]
  \bf{Mihai Anitescu}\footnotemark[2] \\
  Mathematics and Computer Science Division, Argonne National Laboratory \\
  Department of Statistics, University of Chicago \\
}

\begin{document}

\maketitle

\vspace{-\baselineskip}
\begin{abstract}
Nonstationary Gaussian process models can capture complex spatially varying
dependence structures in spatial datasets. However, the large number of
observations in modern datasets makes fitting such models computationally
intractable with conventional dense linear algebra. In addition, derivative-free
or even first-order optimization methods can be slow to converge when estimating
many spatially varying parameters. We present here a computational framework
that couples an algebraic block-diagonal plus low-rank covariance matrix
approximation with stochastic trace estimation to facilitate the efficient use
of second-order solvers for maximum likelihood estimation of Gaussian process
models with many parameters. We demonstrate the effectiveness of these methods
by simultaneously fitting $192$ parameters in the popular nonstationary model of
Paciorek and Schervish using $107{,}600$ sea surface temperature anomaly
measurements.
\end{abstract}

\noindent%
\vfill

\newpage

\section{Introduction} \label{sec:introduction}
Gaussian processes are a prevalent class of models in spatial and spatiotemporal
statistics. This is due in part to the fact that the model is completely
specified by the first two moments of the Gaussian distribution, so that a
practitioner need select only a mean function and covariance function. Let
$Z(\x)$ be a Gaussian process with mean function $\mu(\x) = 0$ and covariance
function
\begin{equation}
  \Cov(Z(\x), Z(\x')) = k(\x, \x' | \prms)
\end{equation}
observed at locations $\{\x_i\}_{i=1}^n$ corresponding to measurements $y_i =
Z(\x_i)$ with $\x_i \in \Omega \subset \R^d$. Here $k(\cdot, \cdot)$ is a
parametric covariance function with parameters $\prms$. Then defining the vector
$\y = [y_1, ..., y_n]^\top \in \R^n$, we have
\begin{equation}
  \y \sim \mathcal{N}\big(\bm{0}, \cm(\prms)\big),
\end{equation}
where the covariance matrix $\cm(\prms) \in \R^{n \times n}$ is defined by
\begin{equation}
  \cm(\prms)_{ij} = \Cov(y_i, y_j) = k(\x_i, \x_j | \prms).
\end{equation}
A primary objective after selecting a parametric covariance model is to estimate
$\prms$ from data. One can then predict the value of the process at unobserved
locations by treating the estimated parameters as the true parameters. A
standard parameter estimation method is to compute the maximum likelihood
estimator (MLE), denoted $\mle$, which minimizes the mean zero Gaussian negative
log-likelihood function
\begin{equation} \label{eq:exactnll}
  -\ell(\prms) = \frac{1}{2}\logdet\cm(\prms) + \frac{1}{2}\y^\top \cm(\prms)^{-1} \y + \frac{n}{2}\log(2\pi).
\end{equation}
For the remainder of this paper we will suppress the dependence of $\cm$ on
$\prms$ for notational clarity. 

Computing (\ref{eq:exactnll}) for large spatial
datasets is computationally challenging, since the determinant and linear solve
operations have cubic time complexity and quadratic space complexity using
conventional dense linear algebra. Thus, for large $n$ evaluating the
log-likelihood directly becomes prohibitively expensive. In response to the
first of these computational challenges, a number of approximations have been
proposed, including Vecchia and ``nearest-neighbor Gaussian process" methods
\cite{vecchia1988estimation, stein2004approximating, katzfuss2017general,
guinness2019gaussian}, matrix tapering \cite{furrer2006covariance}, and Markov
random field approximations \cite{lindgren2011explicit}, which each approximate
$\cm$ or $\cm^{-1}$ using a sparse or block-sparse matrix. Low-rank updates to a
diagonal matrix have also been considered
\cite{cressie2008fixed,banerjee2008gaussian,eidsvik2012approximate,katzfuss2012bayesian,solin2020hilbert},
although they suffer from limitations for nonsmooth fields or when a nugget is
not an appropriate modeling assumption \cite{stein2014limitations}. Approaches
using hierarchical matrices have also been proposed \cite{borm2007approximating,
ambikasaran2015fast, chen2021linear, minden2017fast, litvinenko2019likelihood},
which use rank-structured matrix arithmetic to efficiently evaluate
(\ref{eq:exactnll}).

In spite of the development of these scalable methods, minimizing the negative
log-likelihood remains challenging depending on the covariance function and its
parameterization. Even for covariance functions with few parameters, for example
the stationary isotropic Mat\'ern covariance
\begin{equation} \label{eq:stat-matern}
  \mathcal{M}_\nu(\norm{\x_i - \x_j}) = \frac{\sigma^2}{\Gamma(\nu) 2^{\nu-1}} \bigg(\frac{2\sqrt{\nu} \norm{\x_i - \x_j}}{\rho}\bigg) \mathcal{K}_\nu \bigg(\frac{2\sqrt{\nu} \norm{\x_i - \x_j}}{\rho}\bigg),
\end{equation}
where $\nu$ is a positive constant smoothness parameter and
$\mathcal{K}_\nu(\cdot)$ is the modified Bessel function of the second kind of
order $\nu$, simultaneously estimating the scale $\sigma^2$ and range $\rho$
parameters poses significant challenge \cite{zhang2004inconsistent}. The
likelihood surface is far from convex, and even first-order methods may become
trapped in nearly flat nonellipsoidal regions of the likelihood surface and fail
to make meaningful progress. While a few relatively new articles attempt to
treat the optimization problem more seriously \cite{guinness2019gaussian,
minden2017fast,geoga2019scalable}, the norm in practice is to use
derivative-free or first-order methods with finite difference derivatives.

For more complicated covariance functions that provide more flexibility and thus
require more parameters, the optimization problem becomes even harder, and
second-order optimization can mean the difference between an optimizer
stagnating and successfully reaching the MLE. The computational problem posed in
this setting is that efficient linear solves and log-determinants with $\cm$ are
no longer sufficient. Derivatives $\pj{\cm}$ need to be computed and applied
scalably; and if the gradient and Fisher matrix are computed directly,
matrix-matrix products of the form $\cm^{-1} \Big(\pj{\cm}\Big)$ need to be
computed efficiently. Even for data sizes in which $\cm$ does not require some
form of approximation, with sufficiently many parameters the computational
burden of the matrix-matrix operations necessary for the gradient alone can be
problematic. In this work we address the problem of second-order optimization
for a covariance function with many parameters, using as our motivating example
a nonstationary spatial model, which we introduce now.

For large spatial datasets it is often unrealistic to assume that process
parameters are constant over the entire domain, in other words that the process
is stationary. Therefore nonstationary covariance functions in which the
parameters vary in space become necessary in order to accurately capture the
dependence structure of the data. One such covariance function that we will use
here is derived by Paciorek and Schervish \cite{paciorek2006spatial} as a
modification of the stationary Mat\'ern covariance (\ref{eq:stat-matern}) and is
frequently used in the nonstationary Gaussian process literature
\cite{li2018efficient, risser2015local,banerjee2008gaussian, sang2012full,
huang2021nonstationary}. We use the following anisotropic version,
\begin{equation}
  \label{eq:PS}
  k(\x_i, \x_j) = \sigma^2 \frac{\abs{\Lam(\x_i)}^{\frac{1}{4}} \abs{\Lam(\x_j)}^{\frac{1}{4}}}{\abs{\frac{\Lam(\x_i) + \Lam(\x_j)}{2}}^{\frac{1}{2}}} \mathcal{M}_\nu \Bigg( \sqrt{(\x_i-\x_j)^\top\bigg( \frac{\Lam(\x_i) + \Lam(\x_j)}{2} \bigg)^{-1} (\x_i-\x_j)} \Bigg),
\end{equation}
where $\Lam(\cdot)$ is a spatially varying function that assigns a positive
definite local anisotropy matrix at each location. In fact, Stein
\cite{anderes2011local} gives an extension of this nonstationary covariance that
allows for spatially varying scale $\sigma(\cdot)$ and smoothness $\nu(\cdot)$
functions. However, jointly estimating range and scale parameters can be
challenging even in the simplest stationary settings
\cite{zhang2004inconsistent}, and robustly computing the derivatives of
$\mathcal{M}_\nu$ in the smoothness parameter $\nu$ is numerically
challenging, although progress on this topic has been made recently
\citep{geoga2022}. Thus we restrict our investigations in this work to models
that have only nonstationary local ranges via the anisotropy parameters.

To estimate the spatially varying function $\Lam(\cdot)$ through maximum
likelihood, one can parameterize it using a set of parameter functions
$\theta(\cdot)$ such as its eigenvalues or elements of its Cholesky factors;
expand these spatially varying functions in a basis 
\begin{align} \label{eq:basis}
  \theta(\x) &= \sum_{i=1}^m \phi_i(\x) \, \theta_i,
\end{align}
where $\phi_i$ are some set of basis functions; and estimate the coefficients
$\theta_i$. We adopt this approach here, using a radial basis function (RBF)
expansion (see Section \ref{sec:numericalresults}). Other parameterizations and
bases can be used, however, and we describe here a framework for
high-dimensional parameter optimization with large data that is essentially
agnostic of these choices.

Since large spatial regions or highly nonstationary random fields require a
large number $m$ of basis functions, fitting the entire global model becomes a
difficult high-dimensional optimization problem for which derivative-free and
even first-order solvers are often ineffective. A number of past works using
this model circumvent this difficulty by fitting parameters only locally and
subsequently smoothing or regressing them into a global model using
$(\ref{eq:basis})$ where the $\phi_i$ are RBFs \cite{li2018efficient,
risser2015local,huang2021nonstationary}. Without considering all the basis
coefficients simultaneously, however, relationships between parameters in
adjacent regions are ignored, and it can be difficult to verify the quality of
the global model fit. The primary goal of our work is to present a computational
framework in which this global model fitting is tractable and to consider its
benefits.

We emphasize here that the two computational challenges of large data size and
large parameter dimension are intimately linked. Large data often require
complex models in order to accurately characterize their dependence structure,
and, conversely, complex models often require large data in order to accurately
estimate their many parameters. Therefore, considering these two problems
together is imperative to developing effective practical methods.

We pursue here work in this vein, introducing methods to control the
computational complexity of second-order global optimization. Further, we
provide and discuss an application to high-resolution sea surface temperature
measurements, fitting a dataset of 107,600 measurements to a model with 192
parameters. The results of this application demonstrate that there may be
substantial gains to model fit by estimating the entire global model jointly.

We first introduce substantial improvements to the likelihood computations
provided by \cite{geoga2019scalable} by showing that the hierarchical matrix
approximation therein can be expressed more simply as the \textit{block
full-scale approximation} \cite{snelson2007local, sang2011covariance}. We then
outline the computations of exact gradients, Hessians, and expected Fisher
information matrices in linear time and storage complexity for the block
full-scale approximation with or without a nugget. In addition, we discuss and
apply an accurate stochastic trace estimation method to estimate the gradient
and full dense expected Fisher matrix using a single pass over the derivative
matrices, which offers significant performance gains in practice. With these
strategies, despite the computational burden of a highly expensive covariance
function, we demonstrate that fitting large datasets with many-parameter models
can be done effectively.

We assume that the mean function of the model is zero everywhere and focus
exclusively on covariance matrices. As in the present application, it is fairly
common to study anomaly fields of climatological variables and treat these as
having mean zero for a mean function specified up to some vector of unknown
linear parameters. The matrix calculus for derivatives with respect to these
parameters is much simpler and poses no computational concern.

\section{Block Full-Scale Approximation} \label{sec:representation}
While a number of past works have approximated the covariance matrix as the sum
of a diagonal matrix and a low-rank matrix
\cite{cressie2008fixed,banerjee2008gaussian,eidsvik2012approximate,katzfuss2012bayesian,solin2020hilbert},
these models often fail to capture short-range behavior of the data. To improve
the model fit, one can add a banded or block diagonal matrix to the low-rank
approximation. We consider here a particular algebraic approximation of this
type referred to as the \textit{partially independent conditional} approximation
by Snelson and Gharamani \cite{snelson2007local} or the \textit{block full-scale
approximation} by Sang et al. \cite{sang2011covariance}. This approximation
allows one to modulate the block size and off-diagonal rank independently to
capture both smooth long-range dependence and rough local dependence while still
allowing computations that scale linearly in the number of observations. 

Let $N = \{1,2,...,n\}$ indicate the index set of all observations, and take $I$
and $J$ to be subsets of $N$. Let $\cm_{IJ}$ denote the submatrix of $\cm$
corresponding to rows $I$ and columns $J$. We start by using a k-d tree to
choose a set of $p$ landmark points $\X_P = \{\x_i\}_{i \in P}$ indexed by $P
\subset N$ that are approximately equispaced over the spatial domain, and we
construct the low-rank Nystr\"om approximation
$\cm_{NP}\cm_{PP}^{-1}\cm_{NP}^\top$.

Next, using the k-d tree, we partition the data into disjoint blocks of
observations at nearby spatial locations. Let block $\ell$ consist of
observations indexed by $B_\ell \subset N$ for $\ell=1,...,m$, where $m$ is the
total number of blocks, and denote the collection of these block index sets as
$B = \{B_1, ..., B_m\}$. For a matrix $\A \in \R^{n \times n},$ define the block
diagonalization operator with block structure $B$ as
\begin{equation}
    \Big[\blkdiag_B(\A)\Big]_{ij} = \begin{cases}
      \A_{ij} & \text{for $i,j \in B_\ell$ for some $\ell$} \\
      0 & \text{otherwise.}
    \end{cases}
\end{equation}
We use this operator to construct a block diagonal correction term that takes
the Nystr\"om approximation to the exact values. This yields our approximate
covariance matrix
\begin{equation} \label{eq:BDLR}
    \acm =
    \cm_{NP}\cm_{PP}^{-1}\cm_{NP}^\top +
    \blkdiag_B\Big(\cm-\cm_{NP}\cm_{PP}^{-1}\cm_{NP}^\top\Big),
\end{equation}
where the block diagonal structure allows us to capture short-range covariances
exactly within disjoint local neighborhoods. Alternatively, one can view this
approximation as a two-level approximation to the covariance function given by
\begin{equation}
  \tilde{k}(\x_i, \x_j) = \begin{cases}
    k(\x_i, \x_j) & \text{for $i,j \in B_\ell$ for some $\ell$} \\
    k(\X_P, \x_i)^\top \ k(\X_P, \X_P)^{-1} \ k(\X_P, \x_j) & \text{otherwise.}
  \end{cases}
\end{equation}
This approximation is equivalent to assuming that observations in different
neighborhoods are conditionally independent given the observations at the
landmark points $\X_P$ \cite{snelson2007local}. Considering this conditional
structure, one can also see this approximation as a special case of the Vecchia
approximation \cite{katzfuss2017general}. 

To leverage recent advances in scalable hierarchical matrix operations and
factorizations, Geoga et al. \cite{geoga2019scalable} used the Nystr\"om
approximation to compress off-diagonal blocks within a hierarchical off-diagonal
low-rank (HODLR) approximation to the covariance matrix that could be assembled
and factorized in quasilinear time and storage complexity. In that work, a set
of $p$ landmark points $\X_P = \{\x_i\}_{i \in P}$ are selected from the data
and indexed by $P \subset N$, and each off-diagonal block is approximated using
the low-rank Nystr\"om scheme $\acm_{IJ} = \cm_{IP}\cm_{PP}^{-1}\cm_{JP}^\top$.
Crucially, the set $P$ of landmark points is fixed and used in every
off-diagonal block. Looking entrywise, this approach induces the two-level
matrix approximation
\begin{equation}
  \acm_{ij} = \begin{cases}
    \cm_{ij} & \text{for $i,j \in B_\ell$ for some $\ell$} \\
    \cm_{Pi}^\top\cm_{PP}^{-1}\cm_{Pj} & \text{otherwise},
  \end{cases}
\end{equation}
where the block diagonal entries are exactly the entries of the full covariance
matrix $\cm$ and the nonleaf entries are given by a low-rank approximation. This
is precisely the two-level approximation (\ref{eq:BDLR}). The simplification
from a Nystr\"om-based hierarchical covariance to a two-level covariance is also
discussed by Chen et al. \cite{chen2017hierarchically}.

\section{Linear Complexity Computations} \label{sec:computations}
Here we discuss how to exploit the two-level covariance approximation described
above to obtain direct linear complexity computations of the likelihood and its
derivatives. We derive complexity estimates in terms of the rank of the
Nystr\"om approximation and the block sizes in the block diagonal term. We can
tune these rank and block size parameters to trade off between approximation
accuracy and computational complexity.

\subsection{Computing the log-likelihood}
In order to perform maximum likelihood estimation, our first concern is to
efficiently compute the negative log-likelihood using our approximate covariance
matrix $\acm$, given by
\begin{equation} \label{eq:nll}
  -\ell(\prms) = \frac{1}{2}\logdet\acm + \frac{1}{2}\y^\top \acm^{-1} \y.
\end{equation}
This requires the computation of the determinant of $\acm$ as well as the linear
solve $\acm^{-1} \y$. For this purpose, one would hope to use the matrix
determinant lemma
\begin{equation} \label{eq:detlemma}
  \det\big(\A + \U\V^\top\big) = \det\big(\bm{I} + \V^\top \A^{-1} \U\big) \det\big(\A\big)
\end{equation}
and the Sherman-Woodbury-Morrison formula
\begin{equation} \label{eq:woodbury}
  (\A + \U\V^\top)^{-1} = \A^{-1} - \A^{-1}\U(\bm{I} + \V^\top \A^{-1} \U)^{-1}\V\A^{-1}
\end{equation}
which take advantage of the low-rank update structure for $\A \in \R^{n \times
n}$ and $\U,\V \in \R^{n \times p}$ to reduce the complexity of these
computations. Note, however, that for the Nystr\"om approximation
(\ref{eq:BDLR}) we have $\acm_{PP} = \cm_{PP}\cm_{PP}^{-1}\cm_{PP}^\top =
\cm_{PP}$. In other words, the Nystr\"om approximation is exact on rows and
columns corresponding to landmark points, so
$\blkdiag_B\big(\cm-\cm_{NP}\cm_{PP}^{-1}\cm_{NP}^\top\big)$ is zero on these
rows and columns. In particular it is not invertible, and thus we cannot use
formulas (\ref{eq:detlemma}) and (\ref{eq:woodbury}). Previous works add a
nugget $\sigma^2I$ to $\acm$, circumventing this issue \cite{snelson2007local,
sang2011covariance}. However, this is not strictly necessary. We can remedy the
rank deficiency without a nugget by defining a matrix $\perm$ which permutes the
landmark indices $P$ to the last $p$ indices. Let $Q = N \setminus P$ denote the
index set of non-Nystr\"om indices, and let $B' = [B_1',...,B_m']$ be a new
collection of block index sets which partition $Q$ and thus contain no Nystr\"om
points. One method of constructing $B'$ is to simply remove any Nystr\"om
indices from $B$, namely $B_\ell' = B_\ell \cap Q$. Then we have
\begin{equation} \label{eq:KernelBDLR}
  \acm
  = \perm^\top \mat{\cm_{QP}\cm_{PP}^{-1}\cm_{QP}^\top +
  \blkdiag_{B'}\Big(\cm_{QQ}-\cm_{QP}\cm_{PP}^{-1}\cm_{QP}^\top\Big) & \cm_{QP} \\
  \cm_{QP}^\top & \cm_{PP}} \perm.
\end{equation}
We note that the Nystr\"om rank $p \ll n$, and thus the $(n-p) \times (n-p)$
upper left block of the permuted matrix contains the vast majority of the
covariance information between observations, and the other blocks are small
dense matrices.

This permuted representation yields efficient and convenient computations, since
the block diagonal correction matrix in the upper left block is now full rank.
For ease of notation, we will write this matrix as 
\begin{equation}
  \D = \blkdiag_{B'}\Big(\cm_{QQ}-\cm_{QP}\cm_{PP}^{-1}\cm_{QP}^\top\Big).
\end{equation}
Returning to the determinant of $\acm$, we see that the block diagonal
correction matrix is the Schur complement of $\cm_{PP}$ in the permuted matrix,
and thus we can compute the determinant of the approximate covariance matrix
using the matrix determinant lemma as
\begin{equation} 
  \det\big(\acm\big)
  = \det(\D) \det(\cm_{PP}).
\end{equation}
Here and for the remainder of this section we will assume for ease of analysis
that the block structure $B'$ consists of equally sized blocks of size $b =
\abs{B_\ell'}$ for all $\ell = 1, ..., m$. We will also assume we have
factorized $\cm_{PP}$ and $\D$ as a precomputation step requiring $\O(p^3 +
nb^2)$ work. Therefore, computing the determinant of $\acm$ requires computing
the determinant of one factorized $p \times p$ matrix and $n/b$ determinants of
factorized $b \times b$ blocks, yielding $\O(n)$ complexity overall. 

The linear solve $\acm^{-1} \y$ can also be computed in a convenient way that
leverages Schur complements. Defining the permuted vector $\z = \perm\y$, we
solve the permuted linear system
\begin{align} \label{eq:vec-solve}
  \mat{\w_1 \\ \w_2} &= \mat{\cm_{QP}\cm_{PP}^{-1}\cm_{QP}^\top +
  \D & \cm_{QP} \\
  \cm_{QP}^\top & \cm_{PP}}^{-1} \mat{\z_1 \\ \z_2} \\
  \w_1 & = \D^{-1} \Big( \z_1 - \cm_{QP}\cm_{PP}^{-1}\z_2 \Big) \nonumber \\
  \w_2 & = \cm_{PP}^{-1} \Big( \z_2 - \cm_{QP}^\top\w_1 \Big) \nonumber \\
  \acm^{-1} \y & = \perm^\top \w, \nonumber
\end{align}
which requires $\O(np + nb)$ work to perform the matrix-vector products and
solve the linear systems. Continuing on to study derivatives of (\ref{eq:nll})
in much the same way, we show that even matrix-matrix products with this rank
structure can be worked with conveniently and efficiently, from which
scalability of all the required linear algebra follows.

\subsection{Computing the gradient}
To employ gradient-based optimization algorithms for maximum likelihood
estimation, we must compute the gradient of the negative log-likelihood
(\ref{eq:exactnll}). Each component of the gradient is
\begin{equation} \label{eq:grad}
\big[-\nabla\ell(\prms)\big]_j = \frac{1}{2}\tr\Bigg[\acm^{-1} \prpj{\acm}\Bigg]
- \frac{1}{2}\y^\top \acm^{-1} \prpj{\acm} \acm^{-1} \y.
\end{equation}
Alternative rank-structured approximations to $\cm$ are often constructed using
early-terminating pivoted factorizations and are thus not differentiable with
respect to the kernel parameters. As a result, one must introduce an additional
approximation to the derivative matrices $\pj{\cm(\prms)}$. For example, the
hierarchical matrix method by Minden et al. \cite{minden2017fast} computes an
independent hierarchical approximation to each $\cm(\prms)^{-1} \pj{\cm(\prms)}$
in order to compute the trace term in the gradient of the log-likelihood. In
contrast, Geoga et al. \cite{geoga2019scalable} use an algebraic hierarchical
covariance matrix approximation for which the derivatives matrices
$\pj{\cm(\prms)}$ can be computed exactly in quasilinear complexity, but they
must use stochastic estimators to compute the trace term efficiently.

The key observation for efficient computation of the gradient for the block
full-scale approximation is that $\pj{\acm}$ has a similar rank structure to
$\acm$. Since the Nystr\"om factors $\cm_{QP}$ and $\cm_{PP}$ are simply
submatrices of the covariance matrix $\cm$, we can compute the derivatives of
these factors using the derivatives of the covariance function. Following basic
matrix differentiation, we have
\begin{alignat}{3} \label{eq:diffnys}
  &\pj{}\Big(\cm_{QP}\cm_{PP}^{-1}\cm_{QP}^\top\Big) && =
  \prpj{\cm_{QP}}\cm_{PP}^{-1}\cm_{QP}^\top &&-
  \cm_{QP}\cm_{PP}^{-1}\prpj{\cm_{PP}}\cm_{PP}^{-1}\cm_{QP}^\top \nonumber \\
  & && &&+ \cm_{QP}\cm_{PP}^{-1}\prpj{\cm_{QP}}^\top.
\end{alignat}
Since the second term on the right side shares a factor with each of the others,
the derivative of the rank-$p$ Nystr\"om approximation has rank at most $2p$.

The derivative of the approximate covariance matrix $\acm$ is then given by
\begin{equation}
  \def\arraystretch{1.5}
\pj{\acm}
 = \perm^\top \mat{
  \displaystyle\pj{}\Big(\cm_{QP}\cm_{PP}^{-1}\cm_{QP}^\top\Big)
  + \pj{\D}
  & \displaystyle\pj{\cm_{QP}} \\
\displaystyle\pj{\cm_{QP}^\top} &
\displaystyle\pj{\cm_{PP}}}
\perm,
\end{equation}
which has the same block rank structure as $\acm$ except that the low-rank
portion of the upper left block has double the rank. 

Given this shared rank structure, we can compute the matrix-matrix linear solve
$\acm^{-1} (\pj{\acm})$ with Schur complements as follows:
\begin{align} \label{eq:mat-solve} 
  \mat{\W_1 & \W_3 \\ \W_2 & \W_4}
  &= \mat{\cm_{QP}\cm_{PP}^{-1}\cm_{QP}^\top +
  \D & \cm_{QP} \\
  \cm_{QP}^\top & \cm_{PP}}^{-1} \mat{
    \displaystyle\pj{}\Big(\cm_{QP}\cm_{PP}^{-1}\cm_{QP}^\top\Big) + \pj{\D} & \pj{\cm_{QP}} \\
    \pj{\cm_{QP}}^\top & \pj{\cm_{PP}}
    } \\
  \W_1 & = \D^{-1}\Bigg( \displaystyle\pj{}\Big(\cm_{QP}\cm_{PP}^{-1}\cm_{QP}^\top\Big) + \pj{\D} - \cm_{QP}\cm_{PP}^{-1}\pj{\cm_{QP}}^\top \Bigg) \nonumber \\
  \W_2 & = \cm_{PP}^{-1}\Bigg( \pj{\cm_{QP}}^\top - \cm_{QP}^\top\W_1 \Bigg) \nonumber \\
  \W_3 & = \D^{-1}\Bigg( \pj{\cm_{QP}} - \cm_{QP}\cm_{PP}^{-1}\pj{\cm_{PP}} \Bigg) \nonumber \\
  \W_4 & = \cm_{PP}^{-1}\Bigg( \pj{\cm_{PP}} - \cm_{QP}^\top\W_3 \Bigg) \nonumber \\
  \acm^{-1} \prpj{\acm} & = \perm \W \perm^\top. \nonumber
\end{align} 
This requires only linear solves and matrix-matrix products involving $n \times
n$ block diagonal matrices, $n \times p$ low-rank factors, and $p \times p$
matrices, resulting in $\O(np^2 + nb^2)$ complexity. In addition, the resulting
matrix $\acm^{-1} (\pj{\acm})$ has the same permuted block diagonal plus
low-rank structure as $\acm$, where $\W_1 \in \R^{(n-p)\times(n-p)}$ is a matrix
of rank at most $2p$ plus a block diagonal term and $\W_2 \in \R^{(n-p)\times
p}$, $\W_3 \in \R^{p \times (n-p)}$, and $\W_4 \in \R^{p\times p}$ are small
dense matrices. The preservation of this rank structure under linear solves will
prove useful for computing second-order information in the next section.

The remaining inner product term in the gradient of the negative log-likelihood
(\ref{eq:grad}) can be computed in $\O(np + nb)$ time by using equation
(\ref{eq:vec-solve}) along with a straightforward block matrix-vector product.
This allows us to compute each entry of the gradient in linear complexity in
$n$.

\subsection{Computing the Fisher matrix}
We can employ the linear solve method above to compute the entries of the
expected Fisher information matrix given by
\begin{equation} \label{eq:fisher}
  \mathcal{I}_{jk} = \frac{1}{2}\tr\Bigg[\acm^{-1} \prpj{\acm} \acm^{-1} \prpk{\acm} \Bigg].
\end{equation}
Computing the terms $\acm^{-1} (\pj{\acm})$ and $\acm^{-1}
(\frac{\partial\acm}{\partial\theta_k})$ using equation (\ref{eq:mat-solve}),
applying a straightforward block matrix-matrix product, and computing the trace
of the resulting rank-structured matrix, we obtain $\O(np^2 + nb^2)$ complexity
per entry. Efficient methods for computing $\mathcal{I}$ facilitate the use of
Fisher scoring algorithms to obtain the MLE and can be used to produce
confidence intervals for estimated parameters. Analogous methods can be used to
compute the Hessian of the negative log-likelihood for use in Newton-based
optimization routines. See the appendix.

\subsection{Symmetrized trace estimation} \label{sec:SAA} Although the above
computations of the gradient and the Fisher and Hessian matrices have the
desired linear scaling in the data size $n$, the matrix-matrix solves and
products in the trace terms require a large number of intermediate allocations,
which are computationally expensive in practice. To provide fast unbiased
estimates of these trace terms, we rely on a sample average approximation based
on the Hutchinson estimator
\cite{hutchinson1989stochastic}
\begin{equation} \label{eq:hutchinson}
  \tr(\A) \approx \frac{1}{s} \sum_{\ell=1}^{s} \u_\ell^\top \A \u_\ell
\end{equation}
with $s$ samples, where $\u_\ell$ are independent symmetric Bernoulli vectors,
although alternatives exist (see \cite{stein2013stochastic}). In particular,
factorizing the approximate covariance as $\acm = \W\W^\top$, one can use the
symmetrized estimator presented by Stein et al. \cite{stein2013stochastic},
which is given by
\begin{equation} \label{eq:stoch-grad}
  \tr\Bigg[\acm^{-1} \prpj{\acm}\Bigg] = \tr\Bigg[\W^{-1} \prpj{\acm} \W^{-\top}\Bigg] \approx \frac{1}{s} \sum_{\ell=1}^s \u_\ell^\top \W^{-1} \prpj{\acm} \W^{-\top} \u_\ell.
\end{equation}
Stein et al. \cite{stein2013stochastic} prove a bound on the variance of this
estimator that is at least as strong as the variance bound on the nonsymmetrized
version, and Geoga et al. \cite{geoga2019scalable} provide numerical results
that indicate greatly improved accuracy with the symmetrized estimator compared
with the standard Hutchinson procedure. 

An analogous symmetrized estimator for the trace terms in the Fisher matrix and
Hessian that require only a small number of linear solves with $\W$ and
matrix-vector products with $\pj{\acm}$ is constructed in
\cite{geoga2019scalable}. The Fisher matrix estimator can be written as
\begin{equation} \label{eq:stoch-fish}
  \I_{jk} \approx \frac{1}{4s} \sum_{\ell=1}^s \u_\ell^\top \W^{-1} \bigg( \pj{\acm} + \pk{\acm} \bigg) \acm^{-1} \bigg( \pj{\acm} + \pk{\acm} \bigg) \W^{-\top} \u_\ell - \frac{1}{2} \I_{jj} - \frac{1}{2} \I_{kk}, 
\end{equation}
where the diagonal terms $\I_{jj}$ and $\I_{kk}$ can be estimated in a trivially
symmetric way. This gives fast stochastic estimators of all quantities necessary
for gradient-based and second-order optimization solvers, which can be computed
in $\O(snp + snb)$ complexity per entry since matrix-vector products with
$\pj{\acm}$ and linear solves with $\W$ require $\O(np + nb)$ work, which we now
show.

\subsection{Symmetric factor computation} \label{sec:symfact} The remaining
concern is to obtain a rank-structured symmetric factor $\W$ for our two-level
approximate covariance matrix $\acm$. The principal challenge is the upper left
block, which we hope to factorize as
\begin{equation}
  \cm_{QP}\cm_{PP}^{-1}\cm_{QP}^\top + \D = \Big( \X\Y^\top + \B \Big) \Big( \X\Y^\top + \B \Big)^\top
\end{equation}
for $\X, \Y \in \R^{(n-p) \times p}$ and $\B$ a block diagonal matrix with the
same block structure $B'$ as $\D$. As discussed in Section
\ref{sec:representation}, our matrix structure is a two-level special case of
the HODLR format; thus we use a single step of the symmetric factorization
algorithm of Ambikasaran et al. \cite{ambikasaran2014fast}, which can be written
concisely by computing Cholesky factors
\begin{align}
  \B\B^\top &= \D \label{eq:sym-fact} \\
  \L\L^\top &= \big(\B^{-1}\cm_{QP}\big)^\top \big(\B^{-1}\cm_{QP}\big) \\
  \M\M^\top &= \id + \L^\top \cm_{PP}^{-1} \L \\
  \X &= \cm_{QP} \\
  \Y &= \L^{-\top}\big(\M - \id\big)\L^{-1}\big(\B^{-1}\cm_{QP}\big)^\top.
\end{align}
This requires only Cholesky factorizations of block diagonal matrices and $p
\times p$ matrices, as well as linear solves and matrix-matrix products
involving $p \times p$ and $(n-p) \times p$ matrices, and thus can be computed
in $\O(np^2 + nb^2)$ time. The symmetric factor $\bm{W}$ is then given by
\begin{equation}
  \W = \perm^\top \mat{\X\Y^\top + \B & 0 \\ \Z^\top & \G} \perm
\end{equation}
where the remaining blocks are defined by
\begin{align}
  \Z &= \Big( \X\Y^\top + \B \Big)^{-1} \cm_{QP} \\
  \G\G^\top &= \cm_{PP} - \Z^\top\Z
\end{align}
with $\Z \in \R^{(n-p) \times p}, \G \in \R^{p \times p}$. Since $\W$ maintains
the same permuted block diagonal plus low-rank structure as $\acm$, we can
compute matrix-vector products and linear solves with $\W$ in  $\O(np + nb)$
time using the Sherman-Woodbury-Morrison formula. This facilitates the SAA
approximations to the gradient, Fisher matrix, and Hessian discussed above and
provides a fast sampling method for the process.

\subsection{Prediction and conditional distributions} 
In addition to fast likelihood computations, the rank structure of $\acm$
facilitates fast kriging and results in a conditional covariance matrix with the
same rank structure as $\acm$. Given a set of locations $\{\x_i^*\}_{i=1}^{n_*}$
indexed by $N_*$ with corresponding process values $y_i^* = Z(\x_i^*)$, we
define the vector $\y_* = [y_1^*,...,y_{n_*}^*]$ and wish to compute the
conditional distribution
\begin{equation}
  \y_* | \y \sim N(\acm_{*}^\top \acm^{-1} \y, \acm_{**} - \acm_{*}^\top\acm^{-1}\acm_{*}).
\end{equation}
where $\big(\acm_{*}\big)_{ij} = \tilde{k}(\x_i, \x^*_j)$ and
$\big(\acm_{**}\big)_{ij} = \tilde{k}(\x^*_i, \x^*_j)$. We assign each point
$\x_i^*$ to a block from the observed data, for example by taking the block with
centroid nearest to $\x_i^*$, resulting in a collection of block index sets $B_*
= [B_1^*, ..., B_m^*]$ that partition $N_*$. To simplify the discussion of
computational complexity, we assume this results in blocks of equal size $b_* =
\abs{B_\ell^*}$. Recalling the permutation $\perm$ that orders the landmark
points to the last indices, we obtain covariance matrices of the form 
\begin{align}
  \acm_{**} &= \cm_{N_*P}\cm_{PP}^{-1}\cm_{N_*P}^\top +
  \blkdiag_{B_*}\Big(\cm_{N_*N_*}-\cm_{N_*P}\cm_{PP}^{-1}\cm_{N_*P}^\top\Big) \\
  \acm_{*} &= \perm^\top \mat{\cm_{QP}\cm_{PP}^{-1}\cm_{N_*P}^\top +
  \blkdiag_{B'B_*}\Big(\cm_{QN_*}-\cm_{QP}\cm_{PP}^{-1}\cm_{N_*P}^\top\Big) \\ \cm_{N_*P}^\top}, \label{eq:cross-cov}
\end{align}
where the nonsymmetric block diagonalization operator is defined by
\begin{equation}
  \Big[\blkdiag_{B'B_*}(\A)\Big]_{ij} = \begin{cases}
    \A_{ij} & \text{for $i \in B'_\ell$ and $j \in B_\ell^*$ for some $\ell$} \\
    0 & \text{otherwise.}
  \end{cases}
\end{equation}
We see that these covariance matrices have rank structures that are minor
variations on the block diagonal plus low-rank structure we have used in the
observed data covariance matrix, its derivatives, and its symmetric factor.

To compute the term $\acm^{-1} \y$ in the conditional mean, we use the linear
solve (\ref{eq:vec-solve}) followed by a straightforward block matrix-vector
product with $\acm_*$, which has complexity $\O(n b_* + np + n_*p)$. 

To compute the term $\acm^{-1} \acm_*$ in the conditional covariance, we use the
first column of the structured matrix-matrix solve (\ref{eq:mat-solve}). This
yields a matrix with the same block structure as (\ref{eq:cross-cov}). We then
compute the block matrix-matrix product with $\acm_*^\top$, which has complexity
$\O(n \, b_*^2 + np^2 + n_* p)$. Importantly, the resulting conditional
covariance matrix is a block diagonal plus a rank-$4p$ matrix. Thus we can
afford to compute and store it, facilitating further computations such as
symmetric factorization using (\ref{eq:sym-fact}) and thus yielding conditional
simulations in linear complexity.

\subsection{Numerical verification of $\O(n)$ complexity}
\begin{figure}[!ht]
  \centering
\begingroup
  \makeatletter
  \providecommand\color[2][]{%
    \GenericError{(gnuplot) \space\space\space\@spaces}{%
      Package color not loaded in conjunction with
      terminal option `colourtext'%
    }{See the gnuplot documentation for explanation.%
    }{Either use 'blacktext' in gnuplot or load the package
      color.sty in LaTeX.}%
    \renewcommand\color[2][]{}%
  }%
  \providecommand\includegraphics[2][]{%
    \GenericError{(gnuplot) \space\space\space\@spaces}{%
      Package graphicx or graphics not loaded%
    }{See the gnuplot documentation for explanation.%
    }{The gnuplot epslatex terminal needs graphicx.sty or graphics.sty.}%
    \renewcommand\includegraphics[2][]{}%
  }%
  \providecommand\rotatebox[2]{#2}%
  \@ifundefined{ifGPcolor}{%
    \newif\ifGPcolor
    \GPcolortrue
  }{}%
  \@ifundefined{ifGPblacktext}{%
    \newif\ifGPblacktext
    \GPblacktexttrue
  }{}%
  \let\gplgaddtomacro\g@addto@macro
  \gdef\gplbacktext{}%
  \gdef\gplfronttext{}%
  \makeatother
  \ifGPblacktext
    \def\colorrgb#1{}%
    \def\colorgray#1{}%
  \else
    \ifGPcolor
      \def\colorrgb#1{\color[rgb]{#1}}%
      \def\colorgray#1{\color[gray]{#1}}%
      \expandafter\def\csname LTw\endcsname{\color{white}}%
      \expandafter\def\csname LTb\endcsname{\color{black}}%
      \expandafter\def\csname LTa\endcsname{\color{black}}%
      \expandafter\def\csname LT0\endcsname{\color[rgb]{1,0,0}}%
      \expandafter\def\csname LT1\endcsname{\color[rgb]{0,1,0}}%
      \expandafter\def\csname LT2\endcsname{\color[rgb]{0,0,1}}%
      \expandafter\def\csname LT3\endcsname{\color[rgb]{1,0,1}}%
      \expandafter\def\csname LT4\endcsname{\color[rgb]{0,1,1}}%
      \expandafter\def\csname LT5\endcsname{\color[rgb]{1,1,0}}%
      \expandafter\def\csname LT6\endcsname{\color[rgb]{0,0,0}}%
      \expandafter\def\csname LT7\endcsname{\color[rgb]{1,0.3,0}}%
      \expandafter\def\csname LT8\endcsname{\color[rgb]{0.5,0.5,0.5}}%
    \else
      \def\colorrgb#1{\color{black}}%
      \def\colorgray#1{\color[gray]{#1}}%
      \expandafter\def\csname LTw\endcsname{\color{white}}%
      \expandafter\def\csname LTb\endcsname{\color{black}}%
      \expandafter\def\csname LTa\endcsname{\color{black}}%
      \expandafter\def\csname LT0\endcsname{\color{black}}%
      \expandafter\def\csname LT1\endcsname{\color{black}}%
      \expandafter\def\csname LT2\endcsname{\color{black}}%
      \expandafter\def\csname LT3\endcsname{\color{black}}%
      \expandafter\def\csname LT4\endcsname{\color{black}}%
      \expandafter\def\csname LT5\endcsname{\color{black}}%
      \expandafter\def\csname LT6\endcsname{\color{black}}%
      \expandafter\def\csname LT7\endcsname{\color{black}}%
      \expandafter\def\csname LT8\endcsname{\color{black}}%
    \fi
  \fi
    \setlength{\unitlength}{0.0500bp}%
    \ifx\gptboxheight\undefined%
      \newlength{\gptboxheight}%
      \newlength{\gptboxwidth}%
      \newsavebox{\gptboxtext}%
    \fi%
    \setlength{\fboxrule}{0.5pt}%
    \setlength{\fboxsep}{1pt}%
    \definecolor{tbcol}{rgb}{1,1,1}%
\begin{picture}(10204.00,2834.00)%
    \gplgaddtomacro\gplbacktext{%
      \csname LTb\endcsname
      \put(378,542){\makebox(0,0)[r]{\strut{}\footnotesize $10^{-4}$}}%
      \put(378,1345){\makebox(0,0)[r]{\strut{}\footnotesize $10^{-2}$}}%
      \put(378,2148){\makebox(0,0)[r]{\strut{}\footnotesize $10^{0}$}}%
      \put(510,-79){\makebox(0,0){\strut{}\footnotesize $2^{9}$}}%
      \put(1004,-79){\makebox(0,0){\strut{}\footnotesize $2^{10}$}}%
      \put(1498,-79){\makebox(0,0){\strut{}\footnotesize $2^{11}$}}%
      \put(1992,-79){\makebox(0,0){\strut{}\footnotesize $2^{12}$}}%
      \put(2487,-79){\makebox(0,0){\strut{}\footnotesize $2^{13}$}}%
      \put(2981,-79){\makebox(0,0){\strut{}\footnotesize $2^{14}$}}%
      \put(3475,-79){\makebox(0,0){\strut{}\footnotesize $2^{15}$}}%
      \put(3969,-79){\makebox(0,0){\strut{}\footnotesize $2^{16}$}}%
      \put(4463,-79){\makebox(0,0){\strut{}\footnotesize $2^{17}$}}%
    }%
    \gplgaddtomacro\gplfronttext{%
      \csname LTb\endcsname
      \put(3740,754){\makebox(0,0)[r]{\strut{}$\acm$}}%
      \csname LTb\endcsname
      \put(3740,534){\makebox(0,0)[r]{\strut{}$\ell$}}%
      \csname LTb\endcsname
      \put(3740,314){\makebox(0,0)[r]{\strut{}$\bm{W}$}}%
    }%
    \gplgaddtomacro\gplbacktext{%
      \csname LTb\endcsname
      \put(5097,141){\makebox(0,0)[r]{\strut{}\footnotesize $10^{-1}$}}%
      \put(5097,944){\makebox(0,0)[r]{\strut{}\footnotesize $10^{0}$}}%
      \put(5097,1746){\makebox(0,0)[r]{\strut{}\footnotesize $10^{1}$}}%
      \put(5097,2549){\makebox(0,0)[r]{\strut{}\footnotesize $10^{2}$}}%
      \put(5229,-79){\makebox(0,0){\strut{}\footnotesize $2^{9}$}}%
      \put(5723,-79){\makebox(0,0){\strut{}\footnotesize $2^{10}$}}%
      \put(6217,-79){\makebox(0,0){\strut{}\footnotesize $2^{11}$}}%
      \put(6711,-79){\makebox(0,0){\strut{}\footnotesize $2^{12}$}}%
      \put(7206,-79){\makebox(0,0){\strut{}\footnotesize $2^{13}$}}%
      \put(7700,-79){\makebox(0,0){\strut{}\footnotesize $2^{14}$}}%
      \put(8194,-79){\makebox(0,0){\strut{}\footnotesize $2^{15}$}}%
      \put(8688,-79){\makebox(0,0){\strut{}\footnotesize $2^{16}$}}%
      \put(9182,-79){\makebox(0,0){\strut{}\footnotesize $2^{17}$}}%
    }%
    \gplgaddtomacro\gplfronttext{%
      \csname LTb\endcsname
      \put(7209,2376){\makebox(0,0)[r]{\strut{}$\nabla \ell$}}%
      \csname LTb\endcsname
      \put(7209,2156){\makebox(0,0)[r]{\strut{}$\mathcal{I}$}}%
      \csname LTb\endcsname
      \put(7209,1936){\makebox(0,0)[r]{\strut{}$\bm{y}_* \, | \, \bm{y}$}}%
    }%
    \gplbacktext
    \put(0,0){\includegraphics[width={510.20bp},height={141.70bp}]{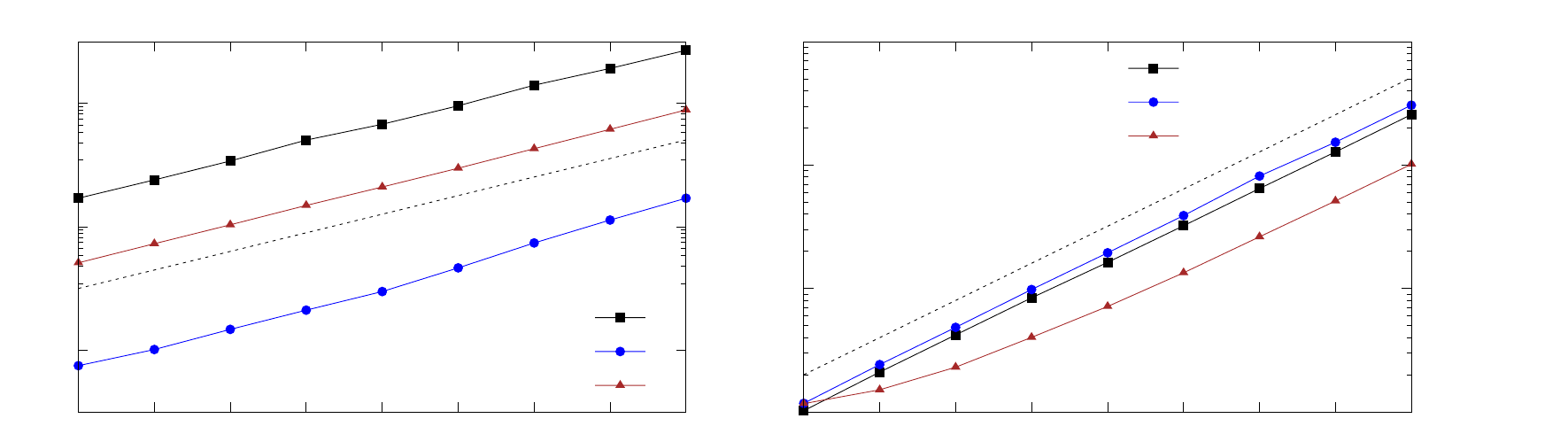}}%
    \gplfronttext
  \end{picture}%
\endgroup

  \caption{Linear scaling of covariance matrix construction, symmetric factorization, and log-likelihood evaluation (left); and of gradient, Fisher matrix, and conditional distribution computation at 512 unobserved locations (right). Dotted black lines show $\O(n)$ scaling for reference.}
  \label{fig:scaling}
\end{figure}

Before applying the methods developed above to a nonstationary process, we
demonstrate their linear scaling using a simple stationary process. We consider
fitting $n$ observations at locations selected uniformly at random in $[0,1]^2$
using the stationary isotropic Mat\'ern covariance (\ref{eq:stat-matern}) with
fixed $\nu=1$, where $\prms = [\sigma^2, \rho]$ are being estimated. We fix the
block size $b=128$ and the Nystr\"om rank $p=32$. For various $n$ we then time
the computation of the approximate covariance matrix $\acm$, the likelihood
$\ell$, the symmetric factor $\W$, the gradient $\nabla \ell$, and the Fisher
information matrix $\I$, as well as computation of the conditional distribution
of $\y_* | \y$ consisting of the mean and rank-structured conditional covariance
matrix at 512 sites also selected uniformly at random in $[0,1]^2$. Figure
\ref{fig:scaling} shows that all the aforementioned computations scale linearly
as expected and gives timings for our implementations on a single core of an
Intel Xeon CPU E5-2650 @ 2.00 GHz machine.

\section{An Application at Scale} \label{sec:numericalresults}
\begin{figure}[!ht]
  \centering
\begingroup
  \makeatletter
  \providecommand\color[2][]{%
    \GenericError{(gnuplot) \space\space\space\@spaces}{%
      Package color not loaded in conjunction with
      terminal option `colourtext'%
    }{See the gnuplot documentation for explanation.%
    }{Either use 'blacktext' in gnuplot or load the package
      color.sty in LaTeX.}%
    \renewcommand\color[2][]{}%
  }%
  \providecommand\includegraphics[2][]{%
    \GenericError{(gnuplot) \space\space\space\@spaces}{%
      Package graphicx or graphics not loaded%
    }{See the gnuplot documentation for explanation.%
    }{The gnuplot epslatex terminal needs graphicx.sty or graphics.sty.}%
    \renewcommand\includegraphics[2][]{}%
  }%
  \providecommand\rotatebox[2]{#2}%
  \@ifundefined{ifGPcolor}{%
    \newif\ifGPcolor
    \GPcolortrue
  }{}%
  \@ifundefined{ifGPblacktext}{%
    \newif\ifGPblacktext
    \GPblacktexttrue
  }{}%
  \let\gplgaddtomacro\g@addto@macro
  \gdef\gplbacktext{}%
  \gdef\gplfronttext{}%
  \makeatother
  \ifGPblacktext
    \def\colorrgb#1{}%
    \def\colorgray#1{}%
  \else
    \ifGPcolor
      \def\colorrgb#1{\color[rgb]{#1}}%
      \def\colorgray#1{\color[gray]{#1}}%
      \expandafter\def\csname LTw\endcsname{\color{white}}%
      \expandafter\def\csname LTb\endcsname{\color{black}}%
      \expandafter\def\csname LTa\endcsname{\color{black}}%
      \expandafter\def\csname LT0\endcsname{\color[rgb]{1,0,0}}%
      \expandafter\def\csname LT1\endcsname{\color[rgb]{0,1,0}}%
      \expandafter\def\csname LT2\endcsname{\color[rgb]{0,0,1}}%
      \expandafter\def\csname LT3\endcsname{\color[rgb]{1,0,1}}%
      \expandafter\def\csname LT4\endcsname{\color[rgb]{0,1,1}}%
      \expandafter\def\csname LT5\endcsname{\color[rgb]{1,1,0}}%
      \expandafter\def\csname LT6\endcsname{\color[rgb]{0,0,0}}%
      \expandafter\def\csname LT7\endcsname{\color[rgb]{1,0.3,0}}%
      \expandafter\def\csname LT8\endcsname{\color[rgb]{0.5,0.5,0.5}}%
    \else
      \def\colorrgb#1{\color{black}}%
      \def\colorgray#1{\color[gray]{#1}}%
      \expandafter\def\csname LTw\endcsname{\color{white}}%
      \expandafter\def\csname LTb\endcsname{\color{black}}%
      \expandafter\def\csname LTa\endcsname{\color{black}}%
      \expandafter\def\csname LT0\endcsname{\color{black}}%
      \expandafter\def\csname LT1\endcsname{\color{black}}%
      \expandafter\def\csname LT2\endcsname{\color{black}}%
      \expandafter\def\csname LT3\endcsname{\color{black}}%
      \expandafter\def\csname LT4\endcsname{\color{black}}%
      \expandafter\def\csname LT5\endcsname{\color{black}}%
      \expandafter\def\csname LT6\endcsname{\color{black}}%
      \expandafter\def\csname LT7\endcsname{\color{black}}%
      \expandafter\def\csname LT8\endcsname{\color{black}}%
    \fi
  \fi
    \setlength{\unitlength}{0.0500bp}%
    \ifx\gptboxheight\undefined%
      \newlength{\gptboxheight}%
      \newlength{\gptboxwidth}%
      \newsavebox{\gptboxtext}%
    \fi%
    \setlength{\fboxrule}{0.5pt}%
    \setlength{\fboxsep}{1pt}%
    \definecolor{tbcol}{rgb}{1,1,1}%
    \scalebox{0.9}{
\begin{picture}(10770.00,6236.00)%
    \gplgaddtomacro\gplbacktext{%
      \csname LTb\endcsname
      \put(406,3561){\makebox(0,0)[r]{\strut{}\footnotesize -15}}%
      \put(406,4028){\makebox(0,0)[r]{\strut{}\footnotesize -7}}%
      \put(406,4496){\makebox(0,0)[r]{\strut{}\footnotesize 1}}%
      \put(406,4964){\makebox(0,0)[r]{\strut{}\footnotesize 9}}%
      \put(406,5431){\makebox(0,0)[r]{\strut{}\footnotesize 17}}%
    }%
    \gplgaddtomacro\gplfronttext{%
      \csname LTb\endcsname
      \put(-67,4442){\rotatebox{-270}{\makebox(0,0){\strut{}latitude}}}%
    }%
    \gplgaddtomacro\gplbacktext{%
    }%
    \gplgaddtomacro\gplfronttext{%
    }%
    \gplgaddtomacro\gplbacktext{%
    }%
    \gplgaddtomacro\gplfronttext{%
      \csname LTb\endcsname
      \put(10012,3273){\makebox(0,0)[l]{\strut{}\footnotesize -5}}%
      \put(10012,3607){\makebox(0,0)[l]{\strut{}\footnotesize -4}}%
      \put(10012,3941){\makebox(0,0)[l]{\strut{}\footnotesize -3}}%
      \put(10012,4275){\makebox(0,0)[l]{\strut{}\footnotesize -2}}%
      \put(10012,4609){\makebox(0,0)[l]{\strut{}\footnotesize -1}}%
      \put(10012,4943){\makebox(0,0)[l]{\strut{}\footnotesize 0}}%
      \put(10012,5277){\makebox(0,0)[l]{\strut{}\footnotesize 1}}%
      \put(10012,5611){\makebox(0,0)[l]{\strut{}\footnotesize 2}}%
    }%
    \gplgaddtomacro\gplbacktext{%
      \csname LTb\endcsname
      \put(406,911){\makebox(0,0)[r]{\strut{}\footnotesize -15}}%
      \put(406,1320){\makebox(0,0)[r]{\strut{}\footnotesize -8}}%
      \put(406,1729){\makebox(0,0)[r]{\strut{}\footnotesize -1}}%
      \put(406,2138){\makebox(0,0)[r]{\strut{}\footnotesize 6}}%
      \put(406,2547){\makebox(0,0)[r]{\strut{}\footnotesize 13}}%
      \put(406,2957){\makebox(0,0)[r]{\strut{}\footnotesize 20}}%
      \put(857,403){\makebox(0,0){\strut{}\footnotesize -115}}%
      \put(1297,403){\makebox(0,0){\strut{}\footnotesize -108}}%
      \put(1736,403){\makebox(0,0){\strut{}\footnotesize -101}}%
      \put(2176,403){\makebox(0,0){\strut{}\footnotesize -94}}%
      \put(2616,403){\makebox(0,0){\strut{}\footnotesize -87}}%
    }%
    \gplgaddtomacro\gplfronttext{%
      \csname LTb\endcsname
      \put(-34,1792){\rotatebox{-270}{\makebox(0,0){\strut{}\footnotesize latitude}}}%
      \put(1794,73){\makebox(0,0){\strut{}\footnotesize longitude}}%
    }%
    \gplgaddtomacro\gplbacktext{%
      \csname LTb\endcsname
      \put(4178,403){\makebox(0,0){\strut{}\footnotesize -115}}%
      \put(4617,403){\makebox(0,0){\strut{}\footnotesize -108}}%
      \put(5057,403){\makebox(0,0){\strut{}\footnotesize -101}}%
      \put(5497,403){\makebox(0,0){\strut{}\footnotesize -94}}%
      \put(5936,403){\makebox(0,0){\strut{}\footnotesize -87}}%
    }%
    \gplgaddtomacro\gplfronttext{%
      \csname LTb\endcsname
      \put(5115,73){\makebox(0,0){\strut{}\footnotesize longitude}}%
      \csname LTb\endcsname
      \put(6691,623){\makebox(0,0)[l]{\strut{}\footnotesize -2}}%
      \put(6691,915){\makebox(0,0)[l]{\strut{}\footnotesize -1.5}}%
      \put(6691,1207){\makebox(0,0)[l]{\strut{}\footnotesize -1}}%
      \put(6691,1499){\makebox(0,0)[l]{\strut{}\footnotesize -0.5}}%
      \put(6691,1792){\makebox(0,0)[l]{\strut{}\footnotesize 0}}%
      \put(6691,2084){\makebox(0,0)[l]{\strut{}\footnotesize 0.5}}%
      \put(6691,2376){\makebox(0,0)[l]{\strut{}\footnotesize 1}}%
      \put(6691,2668){\makebox(0,0)[l]{\strut{}\footnotesize 1.5}}%
      \put(6691,2961){\makebox(0,0)[l]{\strut{}\footnotesize 2}}%
    }%
    \gplgaddtomacro\gplbacktext{%
      \csname LTb\endcsname
      \put(7499,403){\makebox(0,0){\strut{}\footnotesize -115}}%
      \put(7938,403){\makebox(0,0){\strut{}\footnotesize -108}}%
      \put(8378,403){\makebox(0,0){\strut{}\footnotesize -101}}%
      \put(8818,403){\makebox(0,0){\strut{}\footnotesize -94}}%
      \put(9257,403){\makebox(0,0){\strut{}\footnotesize -87}}%
    }%
    \gplgaddtomacro\gplfronttext{%
      \csname LTb\endcsname
      \put(8436,73){\makebox(0,0){\strut{}\footnotesize longitude}}%
      \csname LTb\endcsname
      \put(10012,623){\makebox(0,0)[l]{\strut{}\footnotesize 0}}%
      \put(10012,1012){\makebox(0,0)[l]{\strut{}\footnotesize 0.05}}%
      \put(10012,1402){\makebox(0,0)[l]{\strut{}\footnotesize 0.1}}%
      \put(10012,1792){\makebox(0,0)[l]{\strut{}\footnotesize 0.15}}%
      \put(10012,2181){\makebox(0,0)[l]{\strut{}\footnotesize 0.2}}%
      \put(10012,2571){\makebox(0,0)[l]{\strut{}\footnotesize 0.25}}%
      \put(10012,2961){\makebox(0,0)[l]{\strut{}\footnotesize 0.3}}%
    }%
    \gplbacktext
    \put(0,0){\includegraphics[width={538.50bp},height={311.80bp}]{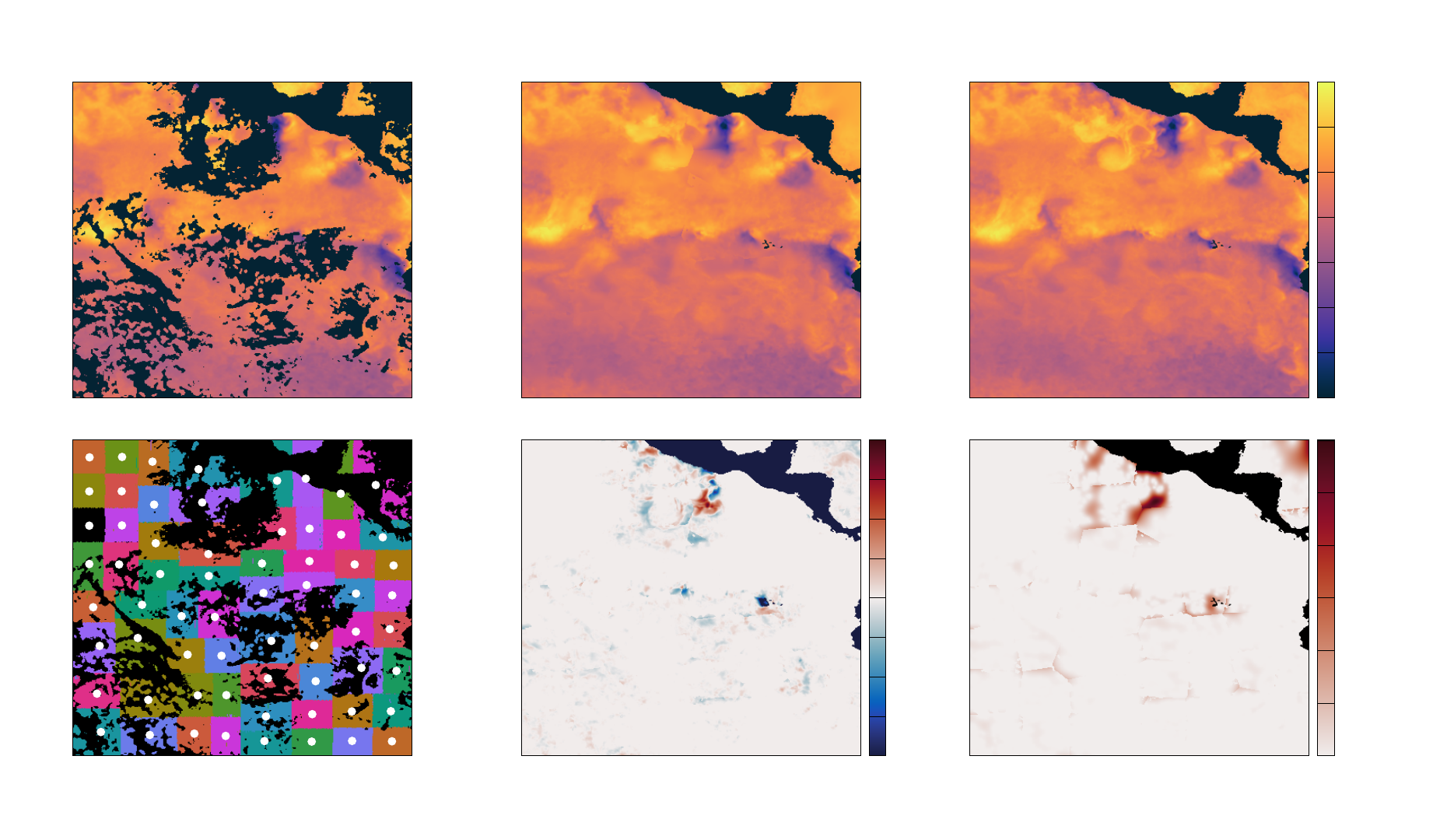}}%
    \gplfronttext
  \end{picture}%
  }
\endgroup
  \caption{Artificially cloud-masked data (top left), interpolant (top center), full data (top right), neighborhoods used for local fitting and approximation blocks with RBF centers as white dots (bottom left), interpolant error (bottom center), and interpolant standard deviations (bottom right).}
  \label{fig:data}
\end{figure}

To test our computational framework in the relevant setting of nonstationary
modeling with many parameters, we study a large sea surface temperature anomaly
dataset from the NOAA Coral Reef Watch database consisting of a 40\textdegree
$\times$ 40\textdegree \ domain in the central Pacific Ocean. These reanalysis
data are de-meaned sea surface temperature measurements generated by
interpolating polar-orbiting and geostationary satellite data from multiple
sources to a $0.05^\circ$ \ ($\approx 5$ km) grid using a Kalman filter-like
approach \cite{maturi2017new, khellah2004statistical}. To study such a large
spatial domain, we subsample this data to a 0.1\textdegree \ grid. We then use
local averaging on NASA's MODIS Cloud Mask product \cite{ackerman2015modis} to
compute a holdout set on the 0.1\textdegree grid, which will serve as a testing
set for evaluating predictions and uncertainties. This leaves 107,600 non-cloudy
observations for maximum likelihood estimation. See Figure \ref{fig:data}. These
cloud-masked data provide a realistic setting for interpolation in atmospheric
science applications.

Recall the nonstationary anisotropic model described in Section
\ref{sec:introduction}, which is given by
\begin{equation}
  \tag{\ref{eq:PS}}
  k(\x_i, \x_j) = \sigma^2 \frac{\abs{\Lam(\x_i)}^{\frac{1}{4}} \abs{\Lam(\x_j)}^{\frac{1}{4}}}{\abs{\frac{\Lam(\x_i) + \Lam(\x_j)}{2}}^{\frac{1}{2}}} \mathcal{M}_\nu \Bigg( \sqrt{(\x_i-\x_j)^\top\bigg( \frac{\Lam(\x_i) + \Lam(\x_j)}{2} \bigg)^{-1} (\x_i-\x_j)} \Bigg).
\end{equation}
To estimate the spatially-varying anisotropy function $\Lam(\cdot)$, we expand
it in a normalized radial basis 
\begin{align} \label{eq:RBF}
  \Lam(\x) &= \sum_{i=1}^m \phi(\norm{\x-\a_i}) \ \Lam_i \\
  \phi(r) &= \frac{\psi(r)}{\sum_{i=1}^m \psi(r)} \\
  \psi(r) &= e^{-(r/c)^2}
\end{align}
using squared exponential bases $\psi$, and we estimate the positive definite
matrices $\Lam_i$. In particular, we express the anisotropy matrices in terms of
their Cholesky factors
\begin{align} \label{eq:aniso}
  \Lam_i &= \bm{L}_i\bm{L}_i^\top \\
  \bm{L}_i &= \mat{\ell_i^{(1,1)} & 0 \\ \ell_i^{(2,1)} & \ell_i^{(2,2)}},
\end{align}
which guarantees positive definiteness. We find also that it is necessary to
parameterize the log of the diagonal elements in order to enforce uniqueness of
the Cholesky factors and avoid identifiability issues. This results in the
parameter vector
\begin{equation} \label{eq:covprms}
  \prms = [\log \ell_1^{(1,1)}, \ell_1^{(2,1)}, \log \ell_1^{(2,2)}, ..., \log \ell_m^{(1,1)}, \ell_m^{(2,1)}, \log \ell_m^{(2,2)}] \in \R^{3m}.
\end{equation}
Since the nonnegative linear combination of positive definite matrices is
positive definite, we have obtained a parameterization in which $\Lam(\x)$ is
positive definite for all $\x \in \Omega$. We fix the smoothness parameter $\nu
= 1$ because it is difficult to estimate simultaneously with the local
anisotropies, and we are interested here primarily in the nonstationary
anisotropy parameters. We estimate the scale parameter $\sigma^2$ using the
\textit{profile likelihood} by fixing $\sigma^2 = 1$, estimating the anisotropy
parameters $\prms$ and then computing the optimal scale parameter, which is
given in closed form by $\sigma^2 = \frac{1}{n} \bm{y}^\top \acm(1, \prms)^{-1}
\bm{y}$.

We start by partitioning our data into 64 disjoint subregions using a k-d tree,
and we use the centroid of each of these regions as the center of an RBF within
the full model, as shown in Figure \ref{fig:data}. To determine a suitable
initial guess for our global model optimization and to compare against the
current state of the art in which locally fitted parameters are plugged into the
RBF expansion (\ref{eq:RBF}), we fit a local anisotropy $\Lam_i$ in each
subregion. Since each subregion contains only 1681 observations, we can afford
to use exact linear algebra and do not require any covariance matrix
approximation. 

To fit these local models as well as to fit the global model we discuss below,
we use our own implementation of the trust-region algorithm adapted directly
from \cite{wright1999numerical}. At iteration $k$ this algorithm minimizes a
quadratic approximation to the negative log-likelihood
\begin{equation} \label{eq:trust-subproblem}
    \min_{\p \in \R^{3m}} -\ell(\prms^{(k)}) - \nabla\ell(\prms^{(k)})^\top \p - \frac{1}{2} \p^\top \B_k \p \quad \text{s.t.} \quad \norm{\p} \leq \Delta_k
\end{equation} 
where $\prms^{(k)}$ are the current parameters, $\B_k$ is an approximation to
the Hessian $\nabla^2 \ell(\prms^{(k)})$, and $\Delta_k$ is a radius parameter
which indicates the size of the region in that this quadratic objective is a
good approximation to the true negative log-likelihood. Trust-region algorithms
are known to converge to stationary points for various approximate solutions of
the subproblem (\ref{eq:trust-subproblem}) and for various Hessian
approximations $\B_k$ as long as $\B_k$ is bounded. In the spirit of Fisher
scoring, we use the symmetrized stochastic estimate of the full Fisher matrix
$\B_k = \I(\prms^{(k)})$ discussed in Section \ref{sec:SAA} and solve the
subproblem (\ref{eq:trust-subproblem}) using a Newton-based iterative method
(\cite{wright1999numerical}, Section 4.3) which is inexpensive and effective for
problems of this size.

Following the local fitting of anisotropy parameters, we optimize the full set
of global model parameters $\prms$. Here we consider the full data and thus
require the block full-scale approximation and accompanying computations
discussed in Sections \ref{sec:representation} and \ref{sec:computations}. We
use the local neighborhoods around each RBF as the blocks in the approximation
and use the k-d tree to select 72 Nystr\"om points for the low-rank portion that
are approximately equispaced over the spatial domain. We can therefore think of
the global model as a combination of the exact local neighborhoods on which the
local parameters were estimated, plus a low-rank term coupling these
neighborhoods. Comparing the likelihood under a disjoint neighborhood model
using the locally estimated anisotropy parameters (i.e. a block diagonal
covariance) with the likelihood under the Paciorek-Schervish model with the
block diagonal plus low-rank covariance approximation using the locally
estimated anisotropy parameters as the $\Lambda_i$ in equation (\ref{eq:RBF}),
we observe a log-likelihood increase of 334{,}342 units. This large likelihood
improvement indicates that observations in adjacent neighborhoods are in fact
highly correlated and that the low-rank Nystr\"om term captures some of this
dependence.

Preparing for optimization of the global model parameters, we choose the width
parameter $c$ of the radial basis function $\psi$ to be half of the minimum
distance between RBF centers. This choice results in bases that are fairly
localized, helping avoid identifiability problems between adjacent RBFs caused
by more diffuse bases with larger $c$ values. In principle, estimating the full
parameter vector $\prms$ allows one to also perform maximum likelihood for the
parameter $c$, which is impossible with local estimation. We find, however, that
if $c$ is included as a parameter alongside the locally estimated parameters,
the derivative in $c$ is much larger than the derivatives in the anisotropy
parameters in $\prms$, and thus the solver moves in the $c$ direction without
materially improving the fit. If we instead fix a localized $c$ and estimate
$\prms$ only, we find that the likelihood surface in $c$ becomes uninformative
because the basis width does not have a large impact on the model fit if the
local parameters are well selected.

\begin{figure}[!ht]
  \centering
\begingroup
  \makeatletter
  \providecommand\color[2][]{%
    \GenericError{(gnuplot) \space\space\space\@spaces}{%
      Package color not loaded in conjunction with
      terminal option `colourtext'%
    }{See the gnuplot documentation for explanation.%
    }{Either use 'blacktext' in gnuplot or load the package
      color.sty in LaTeX.}%
    \renewcommand\color[2][]{}%
  }%
  \providecommand\includegraphics[2][]{%
    \GenericError{(gnuplot) \space\space\space\@spaces}{%
      Package graphicx or graphics not loaded%
    }{See the gnuplot documentation for explanation.%
    }{The gnuplot epslatex terminal needs graphicx.sty or graphics.sty.}%
    \renewcommand\includegraphics[2][]{}%
  }%
  \providecommand\rotatebox[2]{#2}%
  \@ifundefined{ifGPcolor}{%
    \newif\ifGPcolor
    \GPcolortrue
  }{}%
  \@ifundefined{ifGPblacktext}{%
    \newif\ifGPblacktext
    \GPblacktexttrue
  }{}%
  \let\gplgaddtomacro\g@addto@macro
  \gdef\gplbacktext{}%
  \gdef\gplfronttext{}%
  \makeatother
  \ifGPblacktext
    \def\colorrgb#1{}%
    \def\colorgray#1{}%
  \else
    \ifGPcolor
      \def\colorrgb#1{\color[rgb]{#1}}%
      \def\colorgray#1{\color[gray]{#1}}%
      \expandafter\def\csname LTw\endcsname{\color{white}}%
      \expandafter\def\csname LTb\endcsname{\color{black}}%
      \expandafter\def\csname LTa\endcsname{\color{black}}%
      \expandafter\def\csname LT0\endcsname{\color[rgb]{1,0,0}}%
      \expandafter\def\csname LT1\endcsname{\color[rgb]{0,1,0}}%
      \expandafter\def\csname LT2\endcsname{\color[rgb]{0,0,1}}%
      \expandafter\def\csname LT3\endcsname{\color[rgb]{1,0,1}}%
      \expandafter\def\csname LT4\endcsname{\color[rgb]{0,1,1}}%
      \expandafter\def\csname LT5\endcsname{\color[rgb]{1,1,0}}%
      \expandafter\def\csname LT6\endcsname{\color[rgb]{0,0,0}}%
      \expandafter\def\csname LT7\endcsname{\color[rgb]{1,0.3,0}}%
      \expandafter\def\csname LT8\endcsname{\color[rgb]{0.5,0.5,0.5}}%
    \else
      \def\colorrgb#1{\color{black}}%
      \def\colorgray#1{\color[gray]{#1}}%
      \expandafter\def\csname LTw\endcsname{\color{white}}%
      \expandafter\def\csname LTb\endcsname{\color{black}}%
      \expandafter\def\csname LTa\endcsname{\color{black}}%
      \expandafter\def\csname LT0\endcsname{\color{black}}%
      \expandafter\def\csname LT1\endcsname{\color{black}}%
      \expandafter\def\csname LT2\endcsname{\color{black}}%
      \expandafter\def\csname LT3\endcsname{\color{black}}%
      \expandafter\def\csname LT4\endcsname{\color{black}}%
      \expandafter\def\csname LT5\endcsname{\color{black}}%
      \expandafter\def\csname LT6\endcsname{\color{black}}%
      \expandafter\def\csname LT7\endcsname{\color{black}}%
      \expandafter\def\csname LT8\endcsname{\color{black}}%
    \fi
  \fi
    \setlength{\unitlength}{0.0500bp}%
    \ifx\gptboxheight\undefined%
      \newlength{\gptboxheight}%
      \newlength{\gptboxwidth}%
      \newsavebox{\gptboxtext}%
    \fi%
    \setlength{\fboxrule}{0.5pt}%
    \setlength{\fboxsep}{1pt}%
    \definecolor{tbcol}{rgb}{1,1,1}%
    \scalebox{0.8}{
\begin{picture}(9636.00,6236.00)%
    \gplgaddtomacro\gplbacktext{%
      \csname LTb\endcsname
      \put(349,3492){\makebox(0,0)[r]{\strut{}\footnotesize -15}}%
      \put(349,3915){\makebox(0,0)[r]{\strut{}\footnotesize -8}}%
      \put(349,4338){\makebox(0,0)[r]{\strut{}\footnotesize -1}}%
      \put(349,4761){\makebox(0,0)[r]{\strut{}\footnotesize 6}}%
      \put(349,5184){\makebox(0,0)[r]{\strut{}\footnotesize 13}}%
      \put(349,5606){\makebox(0,0)[r]{\strut{}\footnotesize 20}}%
    }%
    \gplgaddtomacro\gplfronttext{%
      \csname LTb\endcsname
      \put(-91,4403){\rotatebox{-270}{\makebox(0,0){\strut{}\footnotesize latitude}}}%
    }%
    \gplgaddtomacro\gplbacktext{%
    }%
    \gplgaddtomacro\gplfronttext{%
    }%
    \gplgaddtomacro\gplbacktext{%
    }%
    \gplgaddtomacro\gplfronttext{%
    }%
    \gplgaddtomacro\gplbacktext{%
      \csname LTb\endcsname
      \put(349,920){\makebox(0,0)[r]{\strut{}\footnotesize -15}}%
      \put(349,1343){\makebox(0,0)[r]{\strut{}\footnotesize -8}}%
      \put(349,1766){\makebox(0,0)[r]{\strut{}\footnotesize -1}}%
      \put(349,2189){\makebox(0,0)[r]{\strut{}\footnotesize 6}}%
      \put(349,2612){\makebox(0,0)[r]{\strut{}\footnotesize 13}}%
      \put(349,3034){\makebox(0,0)[r]{\strut{}\footnotesize 20}}%
      \put(827,403){\makebox(0,0){\strut{}\footnotesize -115}}%
      \put(1305,403){\makebox(0,0){\strut{}\footnotesize -108}}%
      \put(1783,403){\makebox(0,0){\strut{}\footnotesize -101}}%
      \put(2261,403){\makebox(0,0){\strut{}\footnotesize -94}}%
      \put(2738,403){\makebox(0,0){\strut{}\footnotesize -87}}%
    }%
    \gplgaddtomacro\gplfronttext{%
      \csname LTb\endcsname
      \put(-91,1831){\rotatebox{-270}{\makebox(0,0){\strut{}\footnotesize latitude}}}%
      \put(1846,73){\makebox(0,0){\strut{}\footnotesize longitude}}%
    }%
    \gplgaddtomacro\gplbacktext{%
      \csname LTb\endcsname
      \put(3798,403){\makebox(0,0){\strut{}\footnotesize -115}}%
      \put(4276,403){\makebox(0,0){\strut{}\footnotesize -108}}%
      \put(4754,403){\makebox(0,0){\strut{}\footnotesize -101}}%
      \put(5232,403){\makebox(0,0){\strut{}\footnotesize -94}}%
      \put(5709,403){\makebox(0,0){\strut{}\footnotesize -87}}%
    }%
    \gplgaddtomacro\gplfronttext{%
      \csname LTb\endcsname
      \put(4817,73){\makebox(0,0){\strut{}\footnotesize longitude}}%
    }%
    \gplgaddtomacro\gplbacktext{%
      \csname LTb\endcsname
      \put(6770,403){\makebox(0,0){\strut{}\footnotesize -115}}%
      \put(7248,403){\makebox(0,0){\strut{}\footnotesize -108}}%
      \put(7725,403){\makebox(0,0){\strut{}\footnotesize -101}}%
      \put(8203,403){\makebox(0,0){\strut{}\footnotesize -94}}%
      \put(8681,403){\makebox(0,0){\strut{}\footnotesize -87}}%
    }%
    \gplgaddtomacro\gplfronttext{%
      \csname LTb\endcsname
      \put(7788,73){\makebox(0,0){\strut{}\footnotesize longitude}}%
    }%
    \gplbacktext
    \put(0,0){\includegraphics[width={481.80bp},height={311.80bp}]{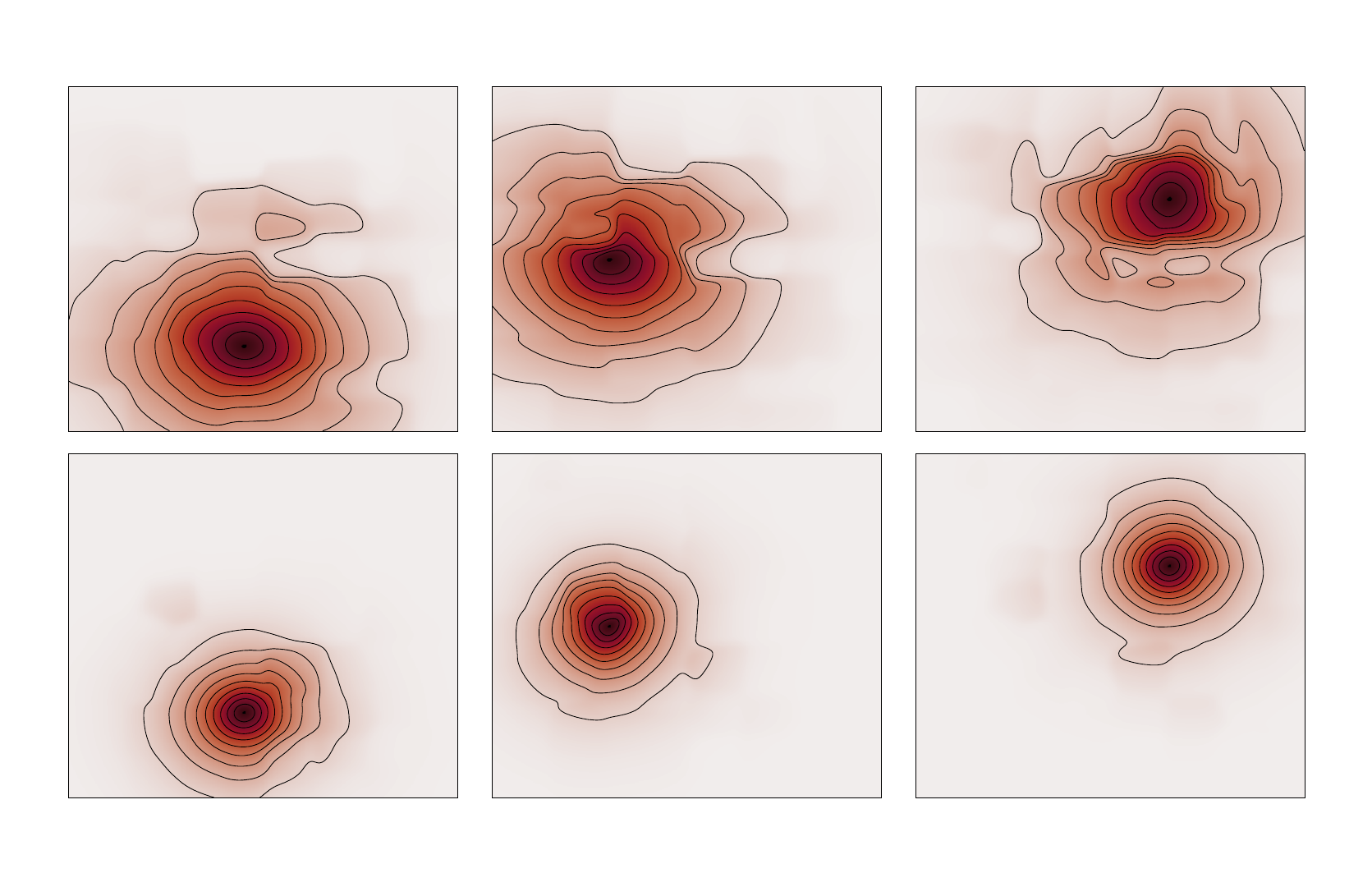}}%
    \gplfronttext
  \end{picture}%
  }
\endgroup
  \caption{Correlation functions at three points using parameters fit globally (top row) and locally (bottom row), with contours at levels $0.1, 0.2, ..., 0.9$.}
  \label{fig:corr}
\end{figure}

After making the block full-scale approximation and selecting the basis width,
the key to fitting these 192 parameters in practice using the trust-region-based
Fisher scoring algorithm described above lies in the use of the SAA methods
discussed in Section \ref{sec:computations} to compute the gradient and Fisher
matrix in a single pass over the derivative matrices $\pj{\acm}$. The crucial
observation is that the symmetrized estimators for both the gradient and Fisher
entries (\ref{eq:stoch-grad}, \ref{eq:stoch-fish}) require only inner products
of terms of the form $\W^{-\top} \u_\ell$ and $\pj{\acm} \W^{-\top} \u_\ell$.
Thus, we can compute each $\pj{\acm}$ and matrix-vector products with
$\W^{-\top} \u_\ell$ efficiently and independently in parallel. We can then
compute the necessary inner products for each entry in the gradient and Fisher
matrix. We use 150 SAA vectors for this numerical experiment, which leads to a
dramatic reduction both in the computational cost, because the inner products
are less expensive than matrix-matrix products, and in memory, because the
vectors $\pj{\acm} \W^{-\top} \u_\ell$ are significantly smaller than the full
derivative matrices $\pj{\acm}$. While both exact and SAA computations require
$\O(n)$ time and storage, SAA provides a dramatic reduction in prefactors. On an
Intel Xeon CPU E5-2650 @ 2.00 GHz machine, computing each Fisher matrix entry
using the exact form takes about 40 seconds in our implementation after the
matrices $\acm^{-1}\pj{\acm}$ have been computed, whereas each entry takes 0.05
seconds using SAA after the vectors $\pj{\acm} \W^{-\top} \u_\ell$ have been
computed. As a result, the full exact Fisher matrix requires about 206 hours,
whereas the SAA approximation requires only 15 minutes. In our implementation,
parallelizing the construction of the derivatives of the covariance matrix
$\pj{\acm}$ on 16 cores leads to approximately 80 seconds for each of the 192
derivative matrices, resulting in a total of about 4 hours. This easily
dominates the computational cost at each iteration, far more demanding than
computing the SAA gradient and Fisher matrix.

Comparing the log-likelihood of the global Paciorek-Schervish model using the
initial, locally fitted parameters with the log-likelihood of the same model
using the final, global parameters given by our Fisher scoring trust-region
algorithm, we observe an increase of 22{,}565 units. In total, parallelized over
16 cores, the time to optimize the local models in disjoint subregions was about
7 hours, and the time to optimize the global model which stagnates after 16
iterations, was about 74 hours. The substantial likelihood difference indicates
that solving the high-dimensional optimization problem to fit all parameters
simultaneously can produce a significant improvement in model fit when compared
with purely local parameter estimation. The difference in covariance structure
can be seen in Figure \ref{fig:corr}, which shows the estimated correlation
function obtained using local parameter estimates versus optimized global
parameter estimates. In particular, we notice that the global model fitting
captures larger east-west correlations above and below the vortices caused by
equatorial currents and little correlation across these currents.

\begin{figure}[!ht]
  \centering
\begingroup
  \makeatletter
  \providecommand\color[2][]{%
    \GenericError{(gnuplot) \space\space\space\@spaces}{%
      Package color not loaded in conjunction with
      terminal option `colourtext'%
    }{See the gnuplot documentation for explanation.%
    }{Either use 'blacktext' in gnuplot or load the package
      color.sty in LaTeX.}%
    \renewcommand\color[2][]{}%
  }%
  \providecommand\includegraphics[2][]{%
    \GenericError{(gnuplot) \space\space\space\@spaces}{%
      Package graphicx or graphics not loaded%
    }{See the gnuplot documentation for explanation.%
    }{The gnuplot epslatex terminal needs graphicx.sty or graphics.sty.}%
    \renewcommand\includegraphics[2][]{}%
  }%
  \providecommand\rotatebox[2]{#2}%
  \@ifundefined{ifGPcolor}{%
    \newif\ifGPcolor
    \GPcolortrue
  }{}%
  \@ifundefined{ifGPblacktext}{%
    \newif\ifGPblacktext
    \GPblacktexttrue
  }{}%
  \let\gplgaddtomacro\g@addto@macro
  \gdef\gplbacktext{}%
  \gdef\gplfronttext{}%
  \makeatother
  \ifGPblacktext
    \def\colorrgb#1{}%
    \def\colorgray#1{}%
  \else
    \ifGPcolor
      \def\colorrgb#1{\color[rgb]{#1}}%
      \def\colorgray#1{\color[gray]{#1}}%
      \expandafter\def\csname LTw\endcsname{\color{white}}%
      \expandafter\def\csname LTb\endcsname{\color{black}}%
      \expandafter\def\csname LTa\endcsname{\color{black}}%
      \expandafter\def\csname LT0\endcsname{\color[rgb]{1,0,0}}%
      \expandafter\def\csname LT1\endcsname{\color[rgb]{0,1,0}}%
      \expandafter\def\csname LT2\endcsname{\color[rgb]{0,0,1}}%
      \expandafter\def\csname LT3\endcsname{\color[rgb]{1,0,1}}%
      \expandafter\def\csname LT4\endcsname{\color[rgb]{0,1,1}}%
      \expandafter\def\csname LT5\endcsname{\color[rgb]{1,1,0}}%
      \expandafter\def\csname LT6\endcsname{\color[rgb]{0,0,0}}%
      \expandafter\def\csname LT7\endcsname{\color[rgb]{1,0.3,0}}%
      \expandafter\def\csname LT8\endcsname{\color[rgb]{0.5,0.5,0.5}}%
    \else
      \def\colorrgb#1{\color{black}}%
      \def\colorgray#1{\color[gray]{#1}}%
      \expandafter\def\csname LTw\endcsname{\color{white}}%
      \expandafter\def\csname LTb\endcsname{\color{black}}%
      \expandafter\def\csname LTa\endcsname{\color{black}}%
      \expandafter\def\csname LT0\endcsname{\color{black}}%
      \expandafter\def\csname LT1\endcsname{\color{black}}%
      \expandafter\def\csname LT2\endcsname{\color{black}}%
      \expandafter\def\csname LT3\endcsname{\color{black}}%
      \expandafter\def\csname LT4\endcsname{\color{black}}%
      \expandafter\def\csname LT5\endcsname{\color{black}}%
      \expandafter\def\csname LT6\endcsname{\color{black}}%
      \expandafter\def\csname LT7\endcsname{\color{black}}%
      \expandafter\def\csname LT8\endcsname{\color{black}}%
    \fi
  \fi
    \setlength{\unitlength}{0.0500bp}%
    \ifx\gptboxheight\undefined%
      \newlength{\gptboxheight}%
      \newlength{\gptboxwidth}%
      \newsavebox{\gptboxtext}%
    \fi%
    \setlength{\fboxrule}{0.5pt}%
    \setlength{\fboxsep}{1pt}%
    \definecolor{tbcol}{rgb}{1,1,1}%
    \scalebox{0.8}{
\begin{picture}(10204.00,3400.00)%
    \gplgaddtomacro\gplbacktext{%
    }%
    \gplgaddtomacro\gplfronttext{%
      \csname LTb\endcsname
      \put(3201,463){\makebox(0,0)[l]{\strut{}\footnotesize -0.3}}%
      \put(3201,875){\makebox(0,0)[l]{\strut{}\footnotesize -0.2}}%
      \put(3201,1287){\makebox(0,0)[l]{\strut{}\footnotesize -0.1}}%
      \put(3201,1699){\makebox(0,0)[l]{\strut{}\footnotesize 0}}%
      \put(3201,2111){\makebox(0,0)[l]{\strut{}\footnotesize 0.1}}%
      \put(3201,2523){\makebox(0,0)[l]{\strut{}\footnotesize 0.2}}%
      \put(3201,2935){\makebox(0,0)[l]{\strut{}\footnotesize 0.3}}%
    }%
    \gplgaddtomacro\gplbacktext{%
    }%
    \gplgaddtomacro\gplfronttext{%
      \csname LTb\endcsname
      \put(6347,463){\makebox(0,0)[l]{\strut{}\footnotesize -0.3}}%
      \put(6347,875){\makebox(0,0)[l]{\strut{}\footnotesize -0.2}}%
      \put(6347,1287){\makebox(0,0)[l]{\strut{}\footnotesize -0.1}}%
      \put(6347,1699){\makebox(0,0)[l]{\strut{}\footnotesize 0}}%
      \put(6347,2111){\makebox(0,0)[l]{\strut{}\footnotesize 0.1}}%
      \put(6347,2523){\makebox(0,0)[l]{\strut{}\footnotesize 0.2}}%
      \put(6347,2935){\makebox(0,0)[l]{\strut{}\footnotesize 0.3}}%
    }%
    \gplgaddtomacro\gplbacktext{%
    }%
    \gplgaddtomacro\gplfronttext{%
      \csname LTb\endcsname
      \put(9493,340){\makebox(0,0)[l]{\strut{}\footnotesize -10}}%
      \put(9493,728){\makebox(0,0)[l]{\strut{}\footnotesize -8}}%
      \put(9493,1116){\makebox(0,0)[l]{\strut{}\footnotesize -6}}%
      \put(9493,1505){\makebox(0,0)[l]{\strut{}\footnotesize -4}}%
      \put(9493,1893){\makebox(0,0)[l]{\strut{}\footnotesize -2}}%
      \put(9493,2282){\makebox(0,0)[l]{\strut{}\footnotesize 0}}%
      \put(9493,2670){\makebox(0,0)[l]{\strut{}\footnotesize 2}}%
      \put(9493,3059){\makebox(0,0)[l]{\strut{}\footnotesize 4}}%
    }%
    \gplbacktext
    \put(0,0){\includegraphics[width={510.20bp},height={170.00bp}]{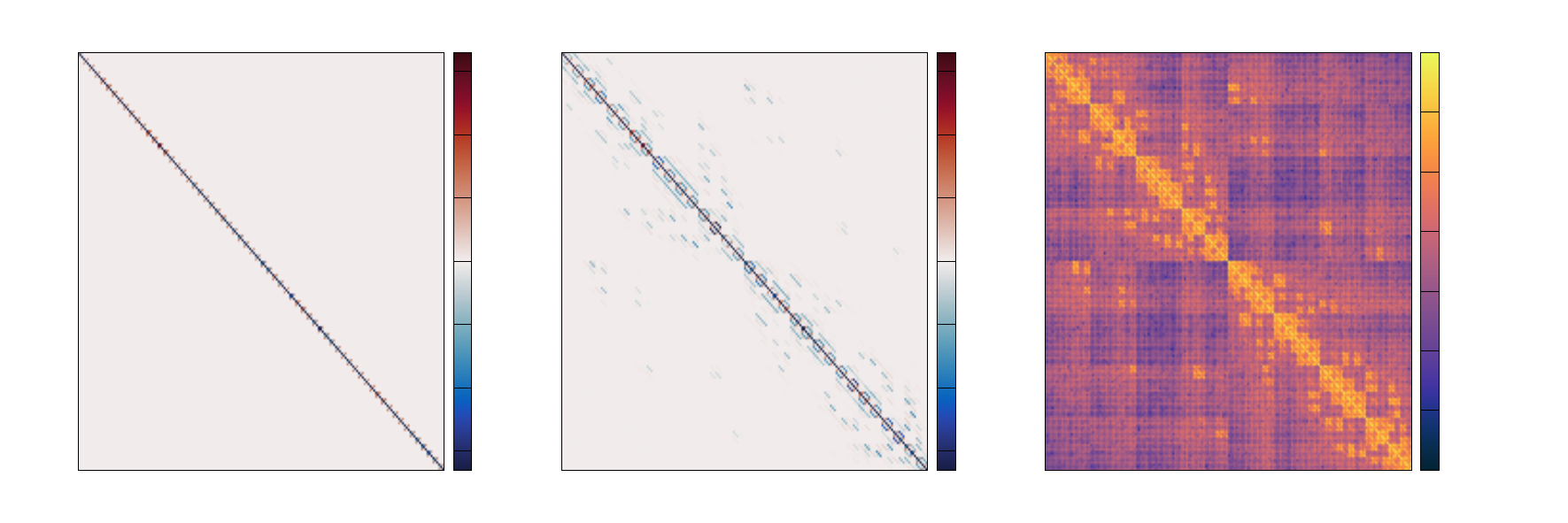}}%
    \gplfronttext
  \end{picture}%
  }
\endgroup
  \caption{Inverse Fisher matrix normalized to have unit diagonal entries for the disjoint local neighborhood model (left) and for the global Paciorek-Schervish model (center) shown with a reduced color range to emphasize off-diagonal elements, as well as the logs of the magnitudes of the entries of the latter Fisher matrix (right).}
  \label{fig:fish}
\end{figure}

\begin{figure}[!ht]
  \centering
\begingroup
  \makeatletter
  \providecommand\color[2][]{%
    \GenericError{(gnuplot) \space\space\space\@spaces}{%
      Package color not loaded in conjunction with
      terminal option `colourtext'%
    }{See the gnuplot documentation for explanation.%
    }{Either use 'blacktext' in gnuplot or load the package
      color.sty in LaTeX.}%
    \renewcommand\color[2][]{}%
  }%
  \providecommand\includegraphics[2][]{%
    \GenericError{(gnuplot) \space\space\space\@spaces}{%
      Package graphicx or graphics not loaded%
    }{See the gnuplot documentation for explanation.%
    }{The gnuplot epslatex terminal needs graphicx.sty or graphics.sty.}%
    \renewcommand\includegraphics[2][]{}%
  }%
  \providecommand\rotatebox[2]{#2}%
  \@ifundefined{ifGPcolor}{%
    \newif\ifGPcolor
    \GPcolortrue
  }{}%
  \@ifundefined{ifGPblacktext}{%
    \newif\ifGPblacktext
    \GPblacktexttrue
  }{}%
  \let\gplgaddtomacro\g@addto@macro
  \gdef\gplbacktext{}%
  \gdef\gplfronttext{}%
  \makeatother
  \ifGPblacktext
    \def\colorrgb#1{}%
    \def\colorgray#1{}%
  \else
    \ifGPcolor
      \def\colorrgb#1{\color[rgb]{#1}}%
      \def\colorgray#1{\color[gray]{#1}}%
      \expandafter\def\csname LTw\endcsname{\color{white}}%
      \expandafter\def\csname LTb\endcsname{\color{black}}%
      \expandafter\def\csname LTa\endcsname{\color{black}}%
      \expandafter\def\csname LT0\endcsname{\color[rgb]{1,0,0}}%
      \expandafter\def\csname LT1\endcsname{\color[rgb]{0,1,0}}%
      \expandafter\def\csname LT2\endcsname{\color[rgb]{0,0,1}}%
      \expandafter\def\csname LT3\endcsname{\color[rgb]{1,0,1}}%
      \expandafter\def\csname LT4\endcsname{\color[rgb]{0,1,1}}%
      \expandafter\def\csname LT5\endcsname{\color[rgb]{1,1,0}}%
      \expandafter\def\csname LT6\endcsname{\color[rgb]{0,0,0}}%
      \expandafter\def\csname LT7\endcsname{\color[rgb]{1,0.3,0}}%
      \expandafter\def\csname LT8\endcsname{\color[rgb]{0.5,0.5,0.5}}%
    \else
      \def\colorrgb#1{\color{black}}%
      \def\colorgray#1{\color[gray]{#1}}%
      \expandafter\def\csname LTw\endcsname{\color{white}}%
      \expandafter\def\csname LTb\endcsname{\color{black}}%
      \expandafter\def\csname LTa\endcsname{\color{black}}%
      \expandafter\def\csname LT0\endcsname{\color{black}}%
      \expandafter\def\csname LT1\endcsname{\color{black}}%
      \expandafter\def\csname LT2\endcsname{\color{black}}%
      \expandafter\def\csname LT3\endcsname{\color{black}}%
      \expandafter\def\csname LT4\endcsname{\color{black}}%
      \expandafter\def\csname LT5\endcsname{\color{black}}%
      \expandafter\def\csname LT6\endcsname{\color{black}}%
      \expandafter\def\csname LT7\endcsname{\color{black}}%
      \expandafter\def\csname LT8\endcsname{\color{black}}%
    \fi
  \fi
    \setlength{\unitlength}{0.0500bp}%
    \ifx\gptboxheight\undefined%
      \newlength{\gptboxheight}%
      \newlength{\gptboxwidth}%
      \newsavebox{\gptboxtext}%
    \fi%
    \setlength{\fboxrule}{0.5pt}%
    \setlength{\fboxsep}{1pt}%
    \definecolor{tbcol}{rgb}{1,1,1}%
    \scalebox{0.8}{
\begin{picture}(10204.00,5668.00)%
    \gplgaddtomacro\gplbacktext{%
      \csname LTb\endcsname
      \put(378,3237){\makebox(0,0)[r]{\strut{}\footnotesize -15}}%
      \put(378,3609){\makebox(0,0)[r]{\strut{}\footnotesize -8}}%
      \put(378,3980){\makebox(0,0)[r]{\strut{}\footnotesize -1}}%
      \put(378,4352){\makebox(0,0)[r]{\strut{}\footnotesize 6}}%
      \put(378,4724){\makebox(0,0)[r]{\strut{}\footnotesize 13}}%
      \put(378,5096){\makebox(0,0)[r]{\strut{}\footnotesize 20}}%
    }%
    \gplgaddtomacro\gplfronttext{%
      \csname LTb\endcsname
      \put(-62,4037){\rotatebox{-270}{\makebox(0,0){\strut{}\footnotesize latitude}}}%
    }%
    \gplgaddtomacro\gplbacktext{%
    }%
    \gplgaddtomacro\gplfronttext{%
    }%
    \gplgaddtomacro\gplbacktext{%
    }%
    \gplgaddtomacro\gplfronttext{%
      \csname LTb\endcsname
      \put(9101,2975){\makebox(0,0)[l]{\strut{}\footnotesize -1}}%
      \put(9101,3506){\makebox(0,0)[l]{\strut{}\footnotesize -0.5}}%
      \put(9101,4037){\makebox(0,0)[l]{\strut{}\footnotesize 0}}%
      \put(9101,4568){\makebox(0,0)[l]{\strut{}\footnotesize 0.5}}%
      \put(9101,5100){\makebox(0,0)[l]{\strut{}\footnotesize 1}}%
    }%
    \gplgaddtomacro\gplbacktext{%
      \csname LTb\endcsname
      \put(378,828){\makebox(0,0)[r]{\strut{}\footnotesize -15}}%
      \put(378,1200){\makebox(0,0)[r]{\strut{}\footnotesize -8}}%
      \put(378,1571){\makebox(0,0)[r]{\strut{}\footnotesize -1}}%
      \put(378,1943){\makebox(0,0)[r]{\strut{}\footnotesize 6}}%
      \put(378,2315){\makebox(0,0)[r]{\strut{}\footnotesize 13}}%
      \put(378,2687){\makebox(0,0)[r]{\strut{}\footnotesize 20}}%
      \put(842,346){\makebox(0,0){\strut{}\footnotesize -115}}%
      \put(1300,346){\makebox(0,0){\strut{}\footnotesize -108}}%
      \put(1758,346){\makebox(0,0){\strut{}\footnotesize -101}}%
      \put(2217,346){\makebox(0,0){\strut{}\footnotesize -94}}%
      \put(2675,346){\makebox(0,0){\strut{}\footnotesize -87}}%
    }%
    \gplgaddtomacro\gplfronttext{%
      \csname LTb\endcsname
      \put(-62,1628){\rotatebox{-270}{\makebox(0,0){\strut{}\footnotesize latitude}}}%
      \put(1819,16){\makebox(0,0){\strut{}\footnotesize longitude}}%
    }%
    \gplgaddtomacro\gplbacktext{%
      \csname LTb\endcsname
      \put(3665,346){\makebox(0,0){\strut{}\footnotesize -115}}%
      \put(4123,346){\makebox(0,0){\strut{}\footnotesize -108}}%
      \put(4581,346){\makebox(0,0){\strut{}\footnotesize -101}}%
      \put(5040,346){\makebox(0,0){\strut{}\footnotesize -94}}%
      \put(5498,346){\makebox(0,0){\strut{}\footnotesize -87}}%
    }%
    \gplgaddtomacro\gplfronttext{%
      \csname LTb\endcsname
      \put(4642,16){\makebox(0,0){\strut{}\footnotesize longitude}}%
    }%
    \gplgaddtomacro\gplbacktext{%
      \csname LTb\endcsname
      \put(6488,346){\makebox(0,0){\strut{}\footnotesize -115}}%
      \put(6946,346){\makebox(0,0){\strut{}\footnotesize -108}}%
      \put(7404,346){\makebox(0,0){\strut{}\footnotesize -101}}%
      \put(7863,346){\makebox(0,0){\strut{}\footnotesize -94}}%
      \put(8321,346){\makebox(0,0){\strut{}\footnotesize -87}}%
    }%
    \gplgaddtomacro\gplfronttext{%
      \csname LTb\endcsname
      \put(7465,16){\makebox(0,0){\strut{}\footnotesize longitude}}%
      \csname LTb\endcsname
      \put(9101,566){\makebox(0,0)[l]{\strut{}\footnotesize -1}}%
      \put(9101,1097){\makebox(0,0)[l]{\strut{}\footnotesize -0.5}}%
      \put(9101,1628){\makebox(0,0)[l]{\strut{}\footnotesize 0}}%
      \put(9101,2159){\makebox(0,0)[l]{\strut{}\footnotesize 0.5}}%
      \put(9101,2691){\makebox(0,0)[l]{\strut{}\footnotesize 1}}%
    }%
    \gplbacktext
    \put(0,0){\includegraphics[width={510.20bp},height={283.40bp}]{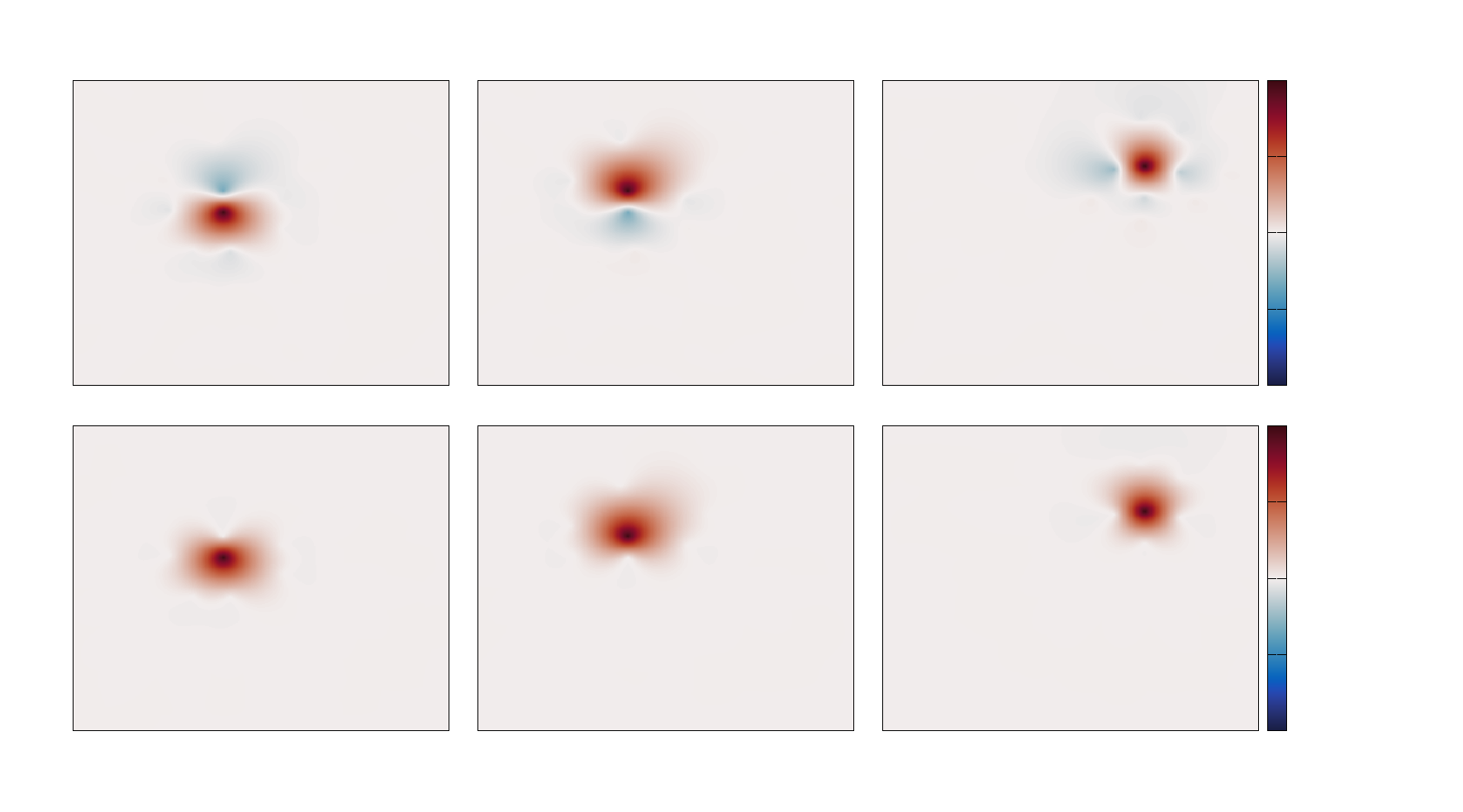}}%
    \gplfronttext
  \end{picture}%
  }
\endgroup
  \caption{Spatially indexed parameter correlation for $\log \ell_i^{(1,1)}$
  using the global Paciorek-Schervish model (top row) and the disjoint local
  neighborhood model (bottom row).}
  \label{fig:param1_corr}
\end{figure}

After maximum likelihood estimation is completed using the inexpensive SAA
gradients and Fisher matrices, we can efficiently compute interpolants and the
rank-structured conditional covariance matrix. Figure \ref{fig:data} shows the
interpolation results along with standard errors. In addition, we compute the
more expensive exact Fisher matrix at the MLE to evaluate parameter
uncertainties. We see that the inverse Fisher matrix contains non-negligible
terms away from the diagonal, indicating that some interaction exists between
spatially proximal parameters. See Figure \ref{fig:fish}, which shows the
correlation matrix given by normalizing the inverse Fisher matrix to have unit
diagonal entries. If one were to fit parameters only locally, the resulting
Fisher matrix would be block diagonal with $3 \times 3$ blocks, which ignores
these off-diagonal contributions (also shown in Figure \ref{fig:fish}). Beyond
the significantly improved likelihoods, this additional second-order information
about parameter point estimates themselves is a material advantage of the global
model formulation used here. Interpolation can be nearly optimal even with a
completely misspecified covariance function \citep{stein1999} and may not be
significantly improved by a global model. We find negligible differences in the
mean squared error of interpolants computed in disjoint local neighborhoods with
stationary models and the interpolants shown in Figure \ref{fig:data} that are
computed with the global nonstationary model. However, uncertainties for
estimated parameters \emph{can} be significantly underestimated when global
dependence is not accounted for. Especially in an RBF model such as the one
presented here, the MLE and expected Fisher matrix from the global model may
also serve as a reasonable approximation to a Bayesian posterior, although one
needs to be careful when appealing to asymptotic results that depend on the
consistency of parameter estimates when working with spatial data
\cite{zhang2004inconsistent}. The difference in parameter correlation structure
can be seen in Figure \ref{fig:param1_corr}, which shows the correlation between
the log of the upper left Cholesky entry $\log \ell_i^{(1,1)}$ (which serves as
one of the nonstationary model parameters, see (\ref{eq:covprms})) at various
spatial locations for the global Paciorek-Schervish model and the disjoint local
neighborhood model. Note the meaningful negative correlations between adjacent
parameters in some regions when using the global model, which cannot be captured
by disjoint local models.

To further identify where significant likelihood improvements have been made, we
investigate quality of fit at the lower tail of the spectrum of the covariance
matrix. If a covariance matrix $\cm \in \mathbb{R}^{n \times n}$ has orthonormal
eigenvectors $\bm{q}_j$ and corresponding positive eigenvalues $\lambda_j$, then
a vector $\bm{z} \sim \mathcal{N}(\bm{0}, \cm)$ can be represented as $\bm{z}
\sim \sum_{j=1}^n \sqrt{\lambda_j} \varepsilon_j \bm{q}_j,$ where $\varepsilon_j
\stackrel{\text{i.i.d.}}{\sim} \mathcal{N}(0,1)$. By virtue of the orthogonality
of the eigenvectors, it follows that under this model $\bm{q}_j^T \bm{z} \sim
\mathcal{N}(0, \lambda_j)$, and $\bm{q}_j^T \bm{z}$ is independent of
$\bm{q}_k^T \bm{z}$ for all $j \neq k$.

\begin{figure}[!ht]
  \centering
\begingroup
  \makeatletter
  \providecommand\color[2][]{%
    \GenericError{(gnuplot) \space\space\space\@spaces}{%
      Package color not loaded in conjunction with
      terminal option `colourtext'%
    }{See the gnuplot documentation for explanation.%
    }{Either use 'blacktext' in gnuplot or load the package
      color.sty in LaTeX.}%
    \renewcommand\color[2][]{}%
  }%
  \providecommand\includegraphics[2][]{%
    \GenericError{(gnuplot) \space\space\space\@spaces}{%
      Package graphicx or graphics not loaded%
    }{See the gnuplot documentation for explanation.%
    }{The gnuplot epslatex terminal needs graphicx.sty or graphics.sty.}%
    \renewcommand\includegraphics[2][]{}%
  }%
  \providecommand\rotatebox[2]{#2}%
  \@ifundefined{ifGPcolor}{%
    \newif\ifGPcolor
    \GPcolortrue
  }{}%
  \@ifundefined{ifGPblacktext}{%
    \newif\ifGPblacktext
    \GPblacktexttrue
  }{}%
  \let\gplgaddtomacro\g@addto@macro
  \gdef\gplbacktext{}%
  \gdef\gplfronttext{}%
  \makeatother
  \ifGPblacktext
    \def\colorrgb#1{}%
    \def\colorgray#1{}%
  \else
    \ifGPcolor
      \def\colorrgb#1{\color[rgb]{#1}}%
      \def\colorgray#1{\color[gray]{#1}}%
      \expandafter\def\csname LTw\endcsname{\color{white}}%
      \expandafter\def\csname LTb\endcsname{\color{black}}%
      \expandafter\def\csname LTa\endcsname{\color{black}}%
      \expandafter\def\csname LT0\endcsname{\color[rgb]{1,0,0}}%
      \expandafter\def\csname LT1\endcsname{\color[rgb]{0,1,0}}%
      \expandafter\def\csname LT2\endcsname{\color[rgb]{0,0,1}}%
      \expandafter\def\csname LT3\endcsname{\color[rgb]{1,0,1}}%
      \expandafter\def\csname LT4\endcsname{\color[rgb]{0,1,1}}%
      \expandafter\def\csname LT5\endcsname{\color[rgb]{1,1,0}}%
      \expandafter\def\csname LT6\endcsname{\color[rgb]{0,0,0}}%
      \expandafter\def\csname LT7\endcsname{\color[rgb]{1,0.3,0}}%
      \expandafter\def\csname LT8\endcsname{\color[rgb]{0.5,0.5,0.5}}%
    \else
      \def\colorrgb#1{\color{black}}%
      \def\colorgray#1{\color[gray]{#1}}%
      \expandafter\def\csname LTw\endcsname{\color{white}}%
      \expandafter\def\csname LTb\endcsname{\color{black}}%
      \expandafter\def\csname LTa\endcsname{\color{black}}%
      \expandafter\def\csname LT0\endcsname{\color{black}}%
      \expandafter\def\csname LT1\endcsname{\color{black}}%
      \expandafter\def\csname LT2\endcsname{\color{black}}%
      \expandafter\def\csname LT3\endcsname{\color{black}}%
      \expandafter\def\csname LT4\endcsname{\color{black}}%
      \expandafter\def\csname LT5\endcsname{\color{black}}%
      \expandafter\def\csname LT6\endcsname{\color{black}}%
      \expandafter\def\csname LT7\endcsname{\color{black}}%
      \expandafter\def\csname LT8\endcsname{\color{black}}%
    \fi
  \fi
    \setlength{\unitlength}{0.0500bp}%
    \ifx\gptboxheight\undefined%
      \newlength{\gptboxheight}%
      \newlength{\gptboxwidth}%
      \newsavebox{\gptboxtext}%
    \fi%
    \setlength{\fboxrule}{0.5pt}%
    \setlength{\fboxsep}{1pt}%
    \definecolor{tbcol}{rgb}{1,1,1}%
    \scalebox{0.9}{
\begin{picture}(9070.00,3400.00)%
    \gplgaddtomacro\gplbacktext{%
      \csname LTb\endcsname
      \put(412,510){\makebox(0,0)[r]{\strut{}\footnotesize 0}}%
      \put(412,800){\makebox(0,0)[r]{\strut{}\footnotesize 0.5}}%
      \put(412,1090){\makebox(0,0)[r]{\strut{}\footnotesize 1}}%
      \put(412,1380){\makebox(0,0)[r]{\strut{}\footnotesize 1.5}}%
      \put(412,1670){\makebox(0,0)[r]{\strut{}\footnotesize 2}}%
      \put(412,1960){\makebox(0,0)[r]{\strut{}\footnotesize 2.5}}%
      \put(412,2250){\makebox(0,0)[r]{\strut{}\footnotesize 3}}%
      \put(412,2541){\makebox(0,0)[r]{\strut{}\footnotesize 3.5}}%
      \put(412,2831){\makebox(0,0)[r]{\strut{}\footnotesize 4}}%
      \put(412,3121){\makebox(0,0)[r]{\strut{}\footnotesize 4.5}}%
      \put(1263,290){\makebox(0,0){\strut{}\footnotesize 0.5}}%
      \put(2001,290){\makebox(0,0){\strut{}\footnotesize 1}}%
      \put(2738,290){\makebox(0,0){\strut{}\footnotesize 1.5}}%
      \put(3476,290){\makebox(0,0){\strut{}\footnotesize 2}}%
      \put(4213,290){\makebox(0,0){\strut{}\footnotesize 2.5}}%
    }%
    \gplgaddtomacro\gplfronttext{%
      \csname LTb\endcsname
      \put(-127,1869){\rotatebox{-270}{\makebox(0,0){\strut{}\footnotesize Empirical quantiles}}}%
      \put(2437,-40){\makebox(0,0){\strut{}\footnotesize Half-normal quantiles}}%
    }%
    \gplgaddtomacro\gplbacktext{%
      \csname LTb\endcsname
      \put(4925,290){\makebox(0,0){\strut{}\footnotesize 0.5}}%
      \put(5306,290){\makebox(0,0){\strut{}\footnotesize 1}}%
      \put(5687,290){\makebox(0,0){\strut{}\footnotesize 1.5}}%
      \put(6069,290){\makebox(0,0){\strut{}\footnotesize 2}}%
      \put(6450,290){\makebox(0,0){\strut{}\footnotesize 2.5}}%
      \put(6831,290){\makebox(0,0){\strut{}\footnotesize 3}}%
      \put(7213,290){\makebox(0,0){\strut{}\footnotesize 3.5}}%
      \put(7594,290){\makebox(0,0){\strut{}\footnotesize 4}}%
      \put(7975,290){\makebox(0,0){\strut{}\footnotesize 4.5}}%
      \put(8475,510){\makebox(0,0)[l]{\strut{}\footnotesize 0}}%
      \put(8475,850){\makebox(0,0)[l]{\strut{}\footnotesize 5}}%
      \put(8475,1190){\makebox(0,0)[l]{\strut{}\footnotesize 10}}%
      \put(8475,1530){\makebox(0,0)[l]{\strut{}\footnotesize 15}}%
      \put(8475,1870){\makebox(0,0)[l]{\strut{}\footnotesize 20}}%
      \put(8475,2209){\makebox(0,0)[l]{\strut{}\footnotesize 25}}%
      \put(8475,2549){\makebox(0,0)[l]{\strut{}\footnotesize 30}}%
      \put(8475,2889){\makebox(0,0)[l]{\strut{}\footnotesize 35}}%
      \put(8475,3229){\makebox(0,0)[l]{\strut{}\footnotesize 40}}%
    }%
    \gplgaddtomacro\gplfronttext{%
      \csname LTb\endcsname
      \put(8915,1869){\rotatebox{-270}{\makebox(0,0){\strut{}\footnotesize count}}}%
      \put(6450,-40){\makebox(0,0){\strut{}\footnotesize $|Z|$}}%
    }%
    \gplbacktext
    \put(0,0){\includegraphics[width={453.50bp},height={170.00bp}]{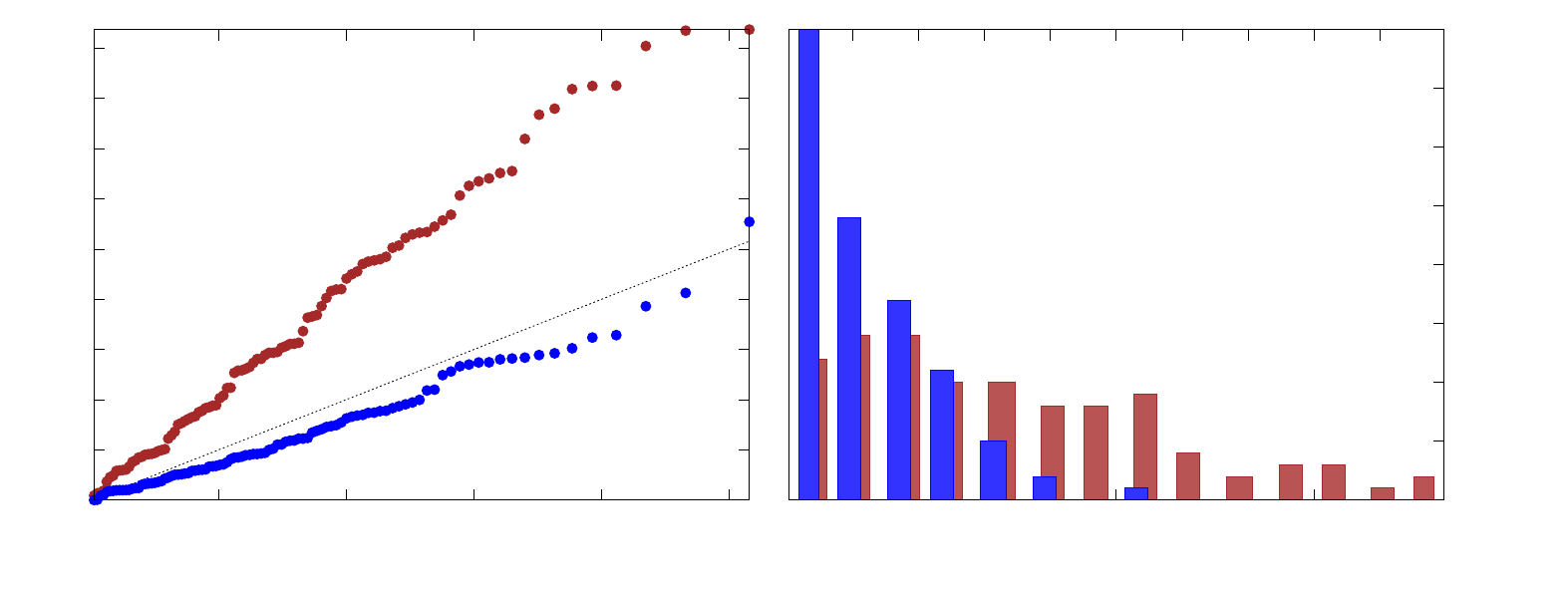}}%
    \gplfronttext
  \end{picture}%
  }
\endgroup
  \caption{Absolute values of standardized $Z$-scores for the eigenvector inner
  products given by equation (\ref{eq:ratio}) corresponding to the 100 smallest
  eigenvalues for the global model covariance matrix using plug-in local
  parameter estimates (red) and globally optimized parameters (blue).}
  \label{fig:eigen_zscore}
\end{figure}

Because $\cm$ is inverted in the quadratic form that appears in the likelihood,
it is particularly revealing of log-likelihood improvements to inspect these
inner products for eigenvectors corresponding to the smallest eigenvalues of
$\cm$.  Figure \ref{fig:eigen_zscore} shows the quantity
\begin{equation} \label{eq:ratio}
  Z_j = \abs{\frac{\bm{q}_j(\prms)^{T}
  \bm{z}}{\sqrt{\lambda_j(\prms)}}},
\end{equation}
the absolute value of a standard $Z$-score, for the eigenpairs corresponding to
the 100 smallest eigenvalues using the model parameters $\prms$ estimated
locally in disjoint regions and simultaneously in the global model.

As can immediately be seen, even after taking into account small-sample
variability, the $Z$-scores using the local plug-in parameters are excessively
dispersed. This reflects a significant misfit, and the corresponding values for
the globally optimized parameters much more closely resemble the standard
half-normal density. Because the smallest eigenvectors for most standard
covariance models generally reflect high-frequency energy and are comparable to
some form of higher-order differencing, this indicates that the global
optimization leads to the model capturing fundamentally local behavior better
than what is achieved with plug-in local parameters. Since the model is
necessarily somewhat misspecified, this behavior is likely explained at least in
part as the global optimization sacrificing fit quality in some parts to make
significant enough improvement in other parts such that the total likelihood is
improved. Considering that almost every model used for serious applications with
environmental data will be to some degree misspecified, we find this to be a
valuable functionality that is not possible with plug-in local parameters.

Overall, we see that the combination of the covariance matrix approximation and
second-order trust-region-based Fisher scoring algorithm presented here, which
scale favorably with the data size and the number of parameters, can produce
meaningfully different fits and richer parameter distributions for
many-parameter models. Therefore these methods represent an important
consideration when modeling nonstationary data at scale.

\section{Discussion} \label{sec:discussion}
In this work we present a complete set of linear cost computations for
second-order maximum likelihood estimation of Gaussian process parameters using
the block full-scale approximation with application to fitting many-parameter
nonstationary models. In particular, we give exact algorithms for computing the
Gaussian log-likelihood and its gradient, Fisher information matrix, and
Hessian, as well as methods for highly efficient stochastic approximation. The
ability to compute derivatives of the log-likelihood is crucial especially in
fitting expressive covariance models with many parameters, in which solvers must
navigate a complex, high-dimensional, nonconvex objective landscape. We
demonstrate using a large sea surface temperature dataset that our methods
facilitate parameter estimation for nonstationary models with a very large
number of parameters. Further, we demonstrate the value of complex global
parameterizations by the significantly improved likelihood and additional
inferred covariance structure between parameters, motivating the need for
methods that are designed to be scalable with respect to both data size and
parameter size.

The block diagonal plus low-rank approximation we use is well known in the
literature as the partially independent conditional \cite{snelson2007local} or
block full-scale approximation \cite{sang2011covariance}. It is also a special
case of the Vecchia approximation \cite{katzfuss2017general} and is a two-level
case of the more general hierarchical models of both Katzfuss
\cite{katzfuss2017multi, katzfuss2020class} and Chen \cite{chen2021linear}.
While these more sophisticated approaches can achieve more accurate
approximations to the log-likelihood \cite{katzfuss2020class, chen2021linear},
they provide no efficient methods for computing its derivatives, making
high-dimensional parameter estimation extremely challenging. Our principal
contributions are to demonstrate that the block full-scale approximation yields
efficient computations with and without a nugget using Schur complements in a
permuted covariance matrix and to apply these computations to high-dimensional
parameter estimation problems. The utility and efficiency of computations with
the block diagonal plus low-rank structure are due to the fact that this
structure is closed with respect to matrix-matrix addition, multiplication, and
inversion. Therefore, by making a single algebraic approximation to the
covariance matrix, we obtain rank-structured derivatives matrices, symmetric
factors, and conditional covariance matrices that can be assembled using only
$\O(n)$ computation and storage.

Although we find that this covariance matrix approximation in conjunction with
the SAA gradient and Fisher matrix methods makes parameter estimation possible
for hundreds of parameters, it has limitations in some circumstances. With
respect to the covariance approximation, Gong et al. \cite{katzfuss2020class}
show that the conditional mean can suffer from discontinuities at block
boundaries, which are noticeable upon close inspection of the interpolant in
Figure \ref{fig:data}. Although these can in theory be alleviated by tapering at
block boundaries with a sufficiently smooth compactly supported function, this
comes at the cost of approximation accuracy. Regarding SAA and parameter
dimension, the full Fisher matrix contains $\O(m^2)$ many entries, where $m$ is
the number of parameters. Due to the incredible efficiency of computing SAA
inner products, this is not a computational bottleneck in the application shown
above. However, if one were to need thousands of parameters with hundreds of
thousands of observations, or if exact Fisher entries were required for
extremely high-precision optimization, this quadratic scaling in parameter
dimension may become prohibitive. In this regime, some form of structured
quasi-Newton that approximates the Fisher matrix using fewer entries may become
necessary. Finally, we note that the cost of evaluating the basis function
expansion (\ref{eq:RBF}) to compute the covariance itself has cost $\O(m)$,
which may become a computational bottleneck for very large datasets with an
extremely large number of basis functions. This could potentially be alleviated
through the use of compactly supported basis functions as in
\cite{huang2021nonstationary}, although these authors use purely local parameter
estimation in place of the global model estimation we have developed here.

\section*{Acknowledgment}
This material was based upon work supported by the US Department of Energy,
Office of Science, Office of Advanced Scientific Computing Research (ASCR) under
Contract DE-AC02-06CH11347. 

\section*{Appendix} \label{sec:appendix}

\subsection*{Computing the Hessian}
To employ second-order Newton solvers instead of Fisher scoring to compute the
MLE, one must compute the Hessian, whose entries are given by
\begin{alignat}{3} \label{eq:hessian} &\big[-\nabla^2\ell(\prms)\big]_{jk} &&=
  -\mathcal{I}_{jk} + \frac{1}{2}\tr\Bigg[\acm^{-1} \prpjk{\acm} \Bigg] &&+
  \y^\top \acm^{-1} \prpj{\acm} \acm^{-1} \prpk{\acm} \acm^{-1} \y \\
  & && &&- \frac{1}{2}\y^\top \acm^{-1} \prpjk{\acm} \acm^{-1} \y. \nonumber
\end{alignat}
We can again apply basic matrix differentiation rules to equation
(\ref{eq:diffnys}) to obtain the second derivatives of the Nystr\"om
approximation, where the second derivative of the rank-$p$ Nystr\"om
approximation
\begin{align}
  \pjk{}\Big(\cm_{QP}\cm_{PP}^{-1}\cm_{QP}^\top\Big) 
  &= \prpjk{\cm_{QP}}\cm_{PP}^{-1}\cm_{QP}^\top \\
  &- \prpj{\cm_{QP}}\cm_{PP}^{-1}\prpk{\cm_{PP}}\cm_{PP}^{-1}\cm_{QP}^\top \nonumber \\
  &+ \prpj{\cm_{QP}}\cm_{PP}^{-1}\prpk{\cm_{QP}^\top} \nonumber \\
  &- \prpk{\cm_{QP}}\cm_{PP}^{-1}\prpj{\cm_{PP}}\cm_{PP}^{-1}\cm_{QP}^\top \nonumber \\ 
  &+ \cm_{QP}\cm_{PP}^{-1}\prpk{\cm_{PP}}\cm_{PP}^{-1}\prpj{\cm_{PP}}\cm_{PP}^{-1}\cm_{QP}^\top \nonumber \\
  &- \cm_{QP}\cm_{PP}^{-1}\prpjk{\cm_{PP}}\cm_{PP}^{-1}\cm_{QP}^\top \nonumber \\
  &+ \cm_{QP}\cm_{PP}^{-1}\prpj{\cm_{PP}}\cm_{PP}^{-1}\prpk{\cm_{PP}}\cm_{PP}^{-1}\cm_{QP}^\top \nonumber \\
  &- \cm_{QP}\cm_{PP}^{-1}\prpj{\cm_{PP}}\cm_{PP}^{-1}\prpk{\cm_{QP}^\top} \nonumber \\
  &+ \prpk{\cm_{QP}}\cm_{PP}^{-1}\prpj{\cm_{QP}}^\top \nonumber \\
  &- \cm_{QP}\cm_{PP}^{-1}\prpk{\cm_{PP}}\cm_{PP}^{-1}\prpj{\cm_{QP}}^\top \nonumber \\
  &+ \cm_{QP}\cm_{PP}^{-1}\prpjk{\cm_{QP}}^\top \nonumber
\end{align}
has rank at most $4p$. The second derivative of the approximate covariance
matrix $\acm$ is then given by
\begin{equation}
  \def\arraystretch{1.5}
\pjk{\acm}
 = \perm^\top \mat{
  \displaystyle\pjk{}\Big(\cm_{QP}\cm_{PP}^{-1}\cm_{QP}^\top\Big)
  + \pjk{\D}
  & \displaystyle\pjk{\cm_{QP}} \\
\displaystyle\pjk{\cm_{QP}^\top} &
\displaystyle\pjk{\cm_{PP}}}
\perm,
\end{equation}
which follows the same permuted block diagonal plus low-rank structure as $\acm$
and $\pj{\acm}$, now with rank at most $4p$ in the low-rank portion of the upper
left block. Thus the trace term in each Hessian entry can be computed using
equation (\ref{eq:mat-solve}) exactly as was done for the analogous term in the
gradient. This results in a linear complexity algorithm for computing entries of
the Hessian.

\bibliography{refs}{}

\begin{thebibliography}{10}

\bibitem{ackerman2015modis}
S.~Ackerman and R.~Frey.
\newblock Modis atmosphere l2 cloud mask product.
\newblock {\em NASA MODIS adaptive processing system, Goddard Space Flight
  Center, USA}, 2015.

\bibitem{ambikasaran2015fast}
S.~Ambikasaran, D.~Foreman-Mackey, L.~Greengard, D.~W. Hogg, and M.~O’Neil.
\newblock Fast direct methods for {G}aussian processes.
\newblock {\em IEEE Transactions on Pattern Analysis and Machine Intelligence},
  38(2):252--265, 2015.

\bibitem{ambikasaran2014fast}
S.~Ambikasaran, M.~O'Neil, and K.~R. Singh.
\newblock Fast symmetric factorization of hierarchical matrices with
  applications.
\newblock {\em arXiv preprint arXiv:1405.0223}, 2014.

\bibitem{anderes2011local}
E.~B. Anderes and M.~L. Stein.
\newblock Local likelihood estimation for nonstationary random fields.
\newblock {\em Journal of Multivariate Analysis}, 102(3):506--520, 2011.

\bibitem{banerjee2008gaussian}
S.~Banerjee, A.~E. Gelfand, A.~O. Finley, and H.~Sang.
\newblock {G}aussian predictive process models for large spatial data sets.
\newblock {\em Journal of the Royal Statistical Society: Series B (Statistical
  Methodology)}, 70(4):825--848, 2008.

\bibitem{borm2007approximating}
S.~B\"orm and J.~Garcke.
\newblock Approximating {G}aussian processes with $\mathcal{H}^2$-matrices.
\newblock In {\em European Conference on Machine Learning}, pages 42--53.
  Springer, 2007.

\bibitem{chen2017hierarchically}
J.~Chen, H.~Avron, and V.~Sindhwani.
\newblock Hierarchically compositional kernels for scalable nonparametric
  learning.
\newblock {\em The Journal of Machine Learning Research}, 18(1):2214--2255,
  2017.

\bibitem{chen2021linear}
J.~Chen and M.~L. Stein.
\newblock Linear-cost covariance functions for {G}aussian random fields.
\newblock {\em Journal of the American Statistical Association}, pages 1--18,
  2021.

\bibitem{cressie2008fixed}
N.~Cressie and G.~Johannesson.
\newblock Fixed rank kriging for very large spatial data sets.
\newblock {\em Journal of the Royal Statistical Society: Series B (Statistical
  Methodology)}, 70(1):209--226, 2008.

\bibitem{eidsvik2012approximate}
J.~Eidsvik, A.~O. Finley, S.~Banerjee, and H.~Rue.
\newblock Approximate {B}ayesian inference for large spatial datasets using
  predictive process models.
\newblock {\em Computational Statistics \& Data Analysis}, 56(6):1362--1380,
  2012.

\bibitem{furrer2006covariance}
R.~Furrer, M.~G. Genton, and D.~Nychka.
\newblock Covariance tapering for interpolation of large spatial datasets.
\newblock {\em Journal of Computational and Graphical Statistics},
  15(3):502--523, 2006.

\bibitem{geoga2019scalable}
C.~J. Geoga, M.~Anitescu, and M.~L. Stein.
\newblock Scalable {G}aussian process computations using hierarchical matrices.
\newblock {\em Journal of Computational and Graphical Statistics}, pages 1--11,
  2019.

\bibitem{geoga2022}
C.~J. Geoga, O.~Marin, M.~Schanen, and M.~L. Stein.
\newblock Fitting {M}at\'ern smoothness parameters using automatic
  differentiation.
\newblock {\em arXiv preprint arXiv:2201.00090}, 2022.

\bibitem{guinness2019gaussian}
J.~Guinness.
\newblock Gaussian process learning via fisher scoring of {V}ecchia's
  approximation.
\newblock {\em Statistics and Computing}, 31(3):1--8, 2021.

\bibitem{huang2021nonstationary}
H.~Huang, L.~R. Blake, M.~Katzfuss, and D.~M. Hammerling.
\newblock Nonstationary spatial modeling of massive global satellite data.
\newblock {\em arXiv preprint arXiv:2111.13428}, 2021.

\bibitem{hutchinson1989stochastic}
M.~F. Hutchinson.
\newblock A stochastic estimator of the trace of the influence matrix for
  {L}aplacian smoothing splines.
\newblock {\em Communications in Statistics-Simulation and Computation},
  18(3):1059--1076, 1989.

\bibitem{katzfuss2017multi}
M.~Katzfuss.
\newblock A multi-resolution approximation for massive spatial datasets.
\newblock {\em Journal of the American Statistical Association},
  112(517):201--214, 2017.

\bibitem{katzfuss2012bayesian}
M.~Katzfuss and N.~Cressie.
\newblock Bayesian hierarchical spatio-temporal smoothing for very large
  datasets.
\newblock {\em Environmetrics}, 23(1):94--107, 2012.

\bibitem{katzfuss2020class}
M.~Katzfuss and W.~Gong.
\newblock A class of multi-resolution approximations for large spatial
  datasets.
\newblock {\em Statistica Sinica}, 30(4):2203--2226, 2020.

\bibitem{katzfuss2017general}
M.~Katzfuss and J.~Guinness.
\newblock A general framework for {V}ecchia approximations of gaussian
  processes.
\newblock {\em Statistical Science}, 36(1):124--141, 2021.

\bibitem{khellah2004statistical}
F.~Khellah, P.~Fieguth, M.~J. Murray, and M.~Allen.
\newblock Statistical processing of large image sequences.
\newblock {\em IEEE Transactions on Image Processing}, 14(1):80--93, 2004.

\bibitem{li2018efficient}
Y.~Li and Y.~Sun.
\newblock Efficient estimation of nonstationary spatial covariance functions
  with application to high-resolution climate model emulation.
\newblock {\em Statistica Sinica}, 29(3):1209--1231, 2019.

\bibitem{lindgren2011explicit}
F.~Lindgren, H.~Rue, and J.~Lindstr{\"o}m.
\newblock An explicit link between {G}aussian fields and {G}aussian markov
  random fields: {T}he stochastic partial differential equation approach.
\newblock {\em Journal of the Royal Statistical Society: Series B (Statistical
  Methodology)}, 73(4):423--498, 2011.

\bibitem{litvinenko2019likelihood}
A.~Litvinenko, Y.~Sun, M.~G. Genton, and D.~E. Keyes.
\newblock Likelihood approximation with hierarchical matrices for large spatial
  datasets.
\newblock {\em Computational Statistics \& Data Analysis}, 137:115--132, 2019.

\bibitem{maturi2017new}
E.~Maturi, A.~Harris, J.~Mittaz, J.~Sapper, G.~Wick, X.~Zhu, P.~Dash, and
  P.~Koner.
\newblock A new high-resolution sea surface temperature blended analysis.
\newblock {\em Bulletin of the American Meteorological Society},
  98(5):1015--1026, 2017.

\bibitem{minden2017fast}
V.~Minden, A.~Damle, K.~L. Ho, and L.~Ying.
\newblock Fast spatial {G}aussian process maximum likelihood estimation via
  skeletonization factorizations.
\newblock {\em Multiscale Modeling \& Simulation}, 15(4):1584--1611, 2017.

\bibitem{paciorek2006spatial}
C.~J. Paciorek and M.~J. Schervish.
\newblock Spatial modelling using a new class of nonstationary covariance
  functions.
\newblock {\em Environmetrics: The official journal of the International
  Environmetrics Society}, 17(5):483--506, 2006.

\bibitem{risser2015local}
M.~D. Risser and C.~A. Calder.
\newblock Local likelihood estimation for covariance functions with
  spatially-varying parameters: the convospat package for r.
\newblock {\em arXiv preprint arXiv:1507.08613}, 2015.

\bibitem{sang2012full}
H.~Sang and J.~Z. Huang.
\newblock A full scale approximation of covariance functions for large spatial
  data sets.
\newblock {\em Journal of the Royal Statistical Society: Series B (Statistical
  Methodology)}, 74(1):111--132, 2012.

\bibitem{sang2011covariance}
H.~Sang, M.~Jun, and J.~Z. Huang.
\newblock Covariance approximation for large multivariate spatial data sets
  with an application to multiple climate model errors.
\newblock {\em The Annals of Applied Statistics}, pages 2519--2548, 2011.

\bibitem{snelson2007local}
E.~Snelson and Z.~Ghahramani.
\newblock Local and global sparse {G}aussian process approximations.
\newblock In {\em Artificial Intelligence and Statistics}, pages 524--531,
  2007.

\bibitem{solin2020hilbert}
A.~Solin and S.~S{\"a}rkk{\"a}.
\newblock Hilbert space methods for reduced-rank {G}aussian process regression.
\newblock {\em Statistics and Computing}, 30(2):419--446, 2020.

\bibitem{stein1999}
M.~L. Stein.
\newblock {\em Interpolation of {Spatial} {Data}: {Some} {Theory} for
  {Kriging}}.
\newblock Springer, 1999.

\bibitem{stein2014limitations}
M.~L. Stein.
\newblock Limitations on low rank approximations for covariance matrices of
  spatial data.
\newblock {\em Spatial Statistics}, 8:1--19, 2014.

\bibitem{stein2013stochastic}
M.~L. Stein, J.~Chen, and M.~Anitescu.
\newblock Stochastic approximation of score functions for {G}aussian processes.
\newblock {\em The Annals of Applied Statistics}, pages 1162--1191, 2013.

\bibitem{stein2004approximating}
M.~L. Stein, Z.~Chi, and L.~J. Welty.
\newblock Approximating likelihoods for large spatial data sets.
\newblock {\em Journal of the Royal Statistical Society: Series B (Statistical
  Methodology)}, 66(2):275--296, 2004.

\bibitem{vecchia1988estimation}
A.V. Vecchia.
\newblock Estimation and model identification for continuous spatial processes.
\newblock {\em Journal of the Royal Statistical Society: Series B (Statistical
  Methodology)}, 50(2):297--312, 1988.

\bibitem{wright1999numerical}
S.~Wright and J.~Nocedal.
\newblock Numerical {O}ptimization.
\newblock {\em Springer Science}, 35(67-68):7, 1999.

\bibitem{zhang2004inconsistent}
H.~Zhang.
\newblock Inconsistent estimation and asymptotically equal interpolations in
  model-based geostatistics.
\newblock {\em Journal of the American Statistical Association},
  99(465):250--261, 2004.

\end{thebibliography}
\bibliographystyle{plain}

\vspace{0.1cm}
\begin{flushright}
  \scriptsize \framebox{\parbox{2.5in}{Government License: The submitted
  manuscript has been created by UChicago Argonne, LLC, Operator of Argonne
  National Laboratory (``Argonne").  Argonne, a US Department of Energy Office
  of Science laboratory, is operated under Contract No. DE-AC02-06CH11357.  The
  US Government retains for itself, and others acting on its behalf, a paid-up
  nonexclusive, irrevocable worldwide license in said article to reproduce,
  prepare derivative works, distribute copies to the public, and perform
  publicly and display publicly, by or on behalf of the Government. The
  Department of Energy will provide public access to these results of federally
  sponsored research in accordance with the DOE Public Access Plan.
  http://energy.gov/downloads/doe-public-access-plan.}} 
  \normalsize
\end{flushright}

\end{document}